\documentclass[12pt]{article}
\usepackage{enumerate}
\usepackage{amsmath}
\usepackage{bm}
\usepackage{tikz}
\begin{document}
\def\be{\begin{equation}}
\def\bea{\begin{eqnarray}}
\def\ee{\end{equation}}
\def\eea{\end{eqnarray}}
\def\d{\partial}
\def\eps{\varepsilon}
\def\la{\lambda}
\def\b{\bigskip}
\def\nn{\nonumber \\}
\def\p{\partial}
\def\t{\tilde}
\def\h{{1\over 2}}
\def\be{\begin{equation}}
\def\bea{\begin{eqnarray}}
\def\ee{\end{equation}}
\def\eea{\end{eqnarray}}
\def\b{\bigskip}
\def\u{\uparrow}
\def \byt{\begin{ytableau}}
\def \eyt{\end{ytableau}}

\overfullrule=0pt
\parskip=2pt
\parindent=12pt
\headheight=0in \headsep=0in \topmargin=0in \oddsidemargin=0in

\vspace{ -3cm}
\thispagestyle{empty}
\vspace{-1cm}

\begin{center}

\vspace{1cm}
{\Large\bf
Separation of variables in the WZW models 
}
\vspace{1.4cm}

{Oleg Lunin$^{a,}$\footnote{olunin@albany.edu  } and Jia Tian$^{b,}$\footnote{wukongjiaozi@ucas.ac.cn } }\\

\vskip 0.5cm

{\em $^{a}$    Department of Physics,
University at Albany (SUNY),
Albany, NY 12222, USA
 }
\ 
\vskip 0.5cm

{\em $^{b}$ Kavli Institute for Theoretical Sciences (KITS),\\
		University of Chinese Academy of Science (UCAS), Beijing 100190, P.~R.~China}
\vskip 0.08cm

\vspace{.2cm}

\end{center}

{\baselineskip 11pt
\begin{abstract}
\noindent
We consider  dynamics of scalar and vector fields on gravitational backgrounds of 
the Wess--Zumino--Witten models. For $SO(4)$ and its cosets, we demonstrate full separation of variables for all fields and find a close analogy with a similar separation of vector equations in the backgrounds of the Myers--Perry black holes. For $SO(5)$ and higher groups separation of variables is found only in some subsectors. 
\end{abstract}
}


\newpage

\newcommand{\comment}[2]{#2}

\tableofcontents
\newpage

\section{Introduction}

Exact solutions of string theory and supergravity provide indispensable insights into dynamics of strongly coupled systems. Once the relevant gravitational backgrounds are found, one extract their physical properties by studying excitations of various fields of these geometries. Unfortunately this problem, which often involves study of partial differential equations (PDEs),
is very difficult unless the geometry has a high degree of symmetry. In such cases one can solve dynamical equations for various fields using separation of variables, and applications of this technique range from simple geometries, such as flat space and spheres, to spacetimes with relatively few isometries, such as rotating black holes in various dimensions. In this article we study separation of variables in the (gauged) Wess--Zumino--Witten (WZW) models \cite{WZW}, the spaces that have either very few isometries or none at all, and show that in certain cases the variables in the scalar and vector equations can be separated. This success is the consequence of the algebraic symmetries of the WZW models, which are guaranteed by the CFT construction, but which are not obvious from the target space perspective. 

The dynamics of a scalar field has been studied in the WZW models and their gauged version, and the full spectrum of eigenvalues is known \cite{PS}\footnote{Similar ideas were also explored earlier in \cite{DVV92}.}. The construction of the relevant wavefunctions is a more complicated problem, and it has been solved only on a case--by--case basis \cite{LTian}. The techniques used in \cite{PS} and \cite{LTian} were purely algebraic, and they did not rely on separation of variables, which is not expected to happen for most (gauged) WZW backgrounds. Unfortunately there are no known procedures for extending these algebraic techniques to vector and tensor fields, especially for finding the eigenfunctions\footnote{Application of algebraic techniques for computing vector eigenvalues will be discussed in \cite{LTianNew}.}. On the other hand, experience with other backgrounds, such as black hole geometries, shows that separation of variables for a scalar field \cite{Carter,Kub1,ChLKil} is often accompanied by separation in the vector and tensor equations \cite{Teuk,LMaxw,OtherMaxw,LHform}. Inspired by this success, we focus on the gWZW models which admit separation of variables in the Helmholtz equation for a scalar and demonstrate that separability of the vector equation in all such cases. We also analyze an example of the WZW model that admits separation of variables only in some subsectors, and we find the explicit expressions for the resulting solutions. 

Wess--Zumino--Witten models are exactly solvable conformal field theories \cite{WZW}, and their numerous applications range from black holes \cite{WittenBH,DVV92} to the quantum Hall effect \cite{QHall}. In the last few years these models have been used to construct families of integrable string theories describing deformations of the systems appearing in the AdS/CFT correspondence \cite{SftsLmb,{HMSLmb},{SuperLmb},{GenLambdaMore}}, and in this context the deformations of the $SO(n)$ and $SO(n)/SO(n-1)$ WZW models are particularly important. Given this interest, we will focus on studying such models with $n=3,4,5$. The eigenvalues and eigenvectors of the scalar field on such background are known through the algebraic constructions \cite{PS,LTian}\footnote{Article \cite{LTian} also solved the scalar equation on the $\la$--deformed backgrounds \cite{SftsLmb}, but we will not discuss such deformations in this paper.}, but it is not clear whether these algebraic methods can be extended to the vector field. By performing an additional analysis of the solutions discussed in \cite{LTian}, we demonstrate an existence of a coordinate system where scalar equations fully separate for the (gauged) WZW models on $SO(4)$ and its cosets, and we identify the Killing--Yano tensors \cite{Yano} associated with this separation. We then study the vector equations on such backgrounds and find the {\it unique} way to separate them. Note that separability of vector equations is more subtle than in the scalar case since knowing the correct coordinate system is not sufficient. One has to identify the correct modes of the vector field as well. In the Myers-Perry geometry of
rotating black holes \cite{MP} this problem was solved in \cite{LMaxw}, where it was shown that the separable components of the gauge field are obtained by taking projections to the frames associated with the Killing--Yano tensor. In this article we prove that the same statement holds for the $SO(4)/H$ WZW models as well, and separation of variables in these cases is very similar to the one found in \cite{LMaxw}. 

Unfortunately the full separation of variable does not seem to persist for the $SO(n)$ WZW model with $n>4$. Nevertheless in the $SO(5)$ several $4$--parametric families of separable solutions which have nontrivial dependences on all $10$ coordinates. These wavefunctions are found by combining algebraic techniques with analysis of PDEs. 

\bigskip

This paper has the following organization. In section \ref{SecSO4} we consider the WZW model for the $SO(4)$ group and demonstrate full separability of dynamical equations for a scalar and a vector. While these results are expected, at least for the scalar, since $SO(4)=SU(2)\times SU(2)$, and a metric on each $SU(2)$ has only one non--cyclic direction, the detailed analysis of separability for the vector field reveals interesting structures which can be extended to the situations where separation of variables is less obvious. We do this in sections \ref{SecSO42}, \ref{SecVecO42}, and \ref{SecSO422}, which focus on the gauged WZW models for various cosets of $SO(4)$. In  section \ref{SecTdual} we show how separable structures on such spaces are mapped into each other under T duality. 

Section \ref{SecSO5} is dedicated to the study of a scalar field on the $SO(5)$ group manifold. While the equations are not fully separable, we identify several important sectors that admit partial separation. Interestingly, all these situations lead to infinite families of solutions which depend on four free parameters. At least one of these families has a simple extension to all $SO(N)$ groups, which is discussed in section \ref{SecSubSpehere}. The results presented in section \ref{SecSO5} are obtained by combining the analysis of the differential equation for the scalar field (sections \ref{SecDressR}, \ref{SecSubXlinear}, \ref{SecSubNoX}, \ref{SecSubSpehere}) and a pure algebraic construction of the eigenfunctions developed in sections \ref{SecSO5group} and \ref{SecSubSymRep}. Some technical details are presented
in the appendices.

\section{Full separation in $SO(4)$ and its cosets}
\label{SecSO4}

The goal of this article is to explore separation of variables for various excitations of the WZW models. In this section we begin with the $SO(4)$ case, where such separation is obvious in the equation for the scalar field. Then we analyze the equations for the vector field and identify the components that separate as well. In sections \ref{SecSO42}--\ref{SecSO422} we  demonstrate that equations for the scalar and vector fields remain separable even after some subgroups of $SO(4)$ are gauged. We show that all these situations follow the pattern discovered in the case of higher dimensional rotating black holes: the separable components of the vector field are constructed by taking projections to the frames associated with the Killing--Yano tensors \cite{LMaxw}. Our analysis serves as a {\it derivation} of the separable ansatz for $SO(4)$ and its cosets since we prove that no other components of the vector fields are separable and that the correct number of polarizations is recovered. 

\subsection{WZW model for $SO(4)$}
\label{SecSubSO4}

We begin with studying excitations of the $SO(4)$ group manifold. The action of the WZW model is given by \cite{WZW}
\bea\label{WZWaction}
S=-\frac{k}{2\pi}\int d^2\sigma \eta^{\alpha\beta}\mbox{tr}(g^{-1}\d_\alpha g g^{-1}\d_\beta g)+
\frac{ik}{6\pi}\int \mbox{tr}(g^{-1}d g\wedge  g^{-1}d g\wedge  g^{-1}d g)\,,
\eea
where $g$ is an element of $SO(4)$. 
Since the WZW background is conformal, the dilaton is trivial\footnote{Gauging of some symmetries leads to a non--trivial dilaton \cite{BarsSfetsDil}. We will discuss this in more detail below.}. We consider various excitations of the background (\ref{WZWaction}), such as scalar and vector fields propagating on the geometry with a metric
\bea\label{MetrSO4a}
ds^2=-\frac{k}{2\pi}\mbox{tr}(g^{-1}d g g^{-1}d g).
\eea
Bearing in mind extensions to larger groups discussed in the next section, we parameterize an element of $SO(4)$ as\footnote{Similar paramaterization for other groups and cosets was introduced in \cite{BarsSfets}.} 
\bea\label{ParamSO4}
g=\begin{bmatrix}
	q_2(\alpha_L)&0\\
	0&{q}_2(\beta_L)
\end{bmatrix} \begin{bmatrix}
	I-\frac{2}{1+XX^T}XX^T&\frac{2}{1+XX^T}{X}\\
	-X^T\frac{2}{1+XX^T}& I-\frac{2}{1+X^TX}X^T X
\end{bmatrix}
\begin{bmatrix}
	q_2(\alpha_R)&0\\
	0&{q}_2(\beta_R)
\end{bmatrix}.
\eea
Here $X=\mbox{diag}(X_1,X_2)$ is a diagonal $2\times 2$ matrix, and $q_2(\gamma)$ are elements of $SO(2)$:
\bea
q_2(\gamma)=\begin{bmatrix}
\cos\gamma&\sin\gamma\\
-\sin\gamma&\cos\gamma
\end{bmatrix}.
\eea 
To justify the parameterization (\ref{ParamSO4}), we observe that the action (\ref{WZWaction}) is invariant under $F=SO(4)\times SO(4)$ transformations, $g\rightarrow h_L g h_R$. To separate variables in various dynamical equations, it is convenient to maximize the number of cyclic directions in the metric (\ref{ParamSO4}). Such cyclic directions correspond to commuting $U(1)$ subgroups of $F$, and there are at most four of them since $F$ has rank four. Therefore, it is convenient to choose a parameterization where the $[U(1)]^4$ Cartan subgroup of $F$ is realized by simple shifts of coordinates $(\alpha_L,\beta_L,\alpha_R,\beta_R)$, and this is accomplished by the introduction of the left and the right matrices in (\ref{ParamSO4}). In sections \ref{SecSO42} and \ref{SecSO422} some elements of the Cartan group 
$[U(1)]^4$ will be gauged by setting some of the four angular coordinates to zero. The matrix in the middle of (\ref{ParamSO4}) contains the remaining two out of six parameters of $SO(4)$. Although one can start with an arbitrary $2\times 2$ matrix 
$X$ there, the transformation $X\rightarrow q_2(\gamma_L) g q_2(\gamma_R)$ can be used to diagonaze that matrix, and parameters $(\gamma_L,\gamma_R)$ can be removed by shifting the 
Cartan coordinates. This leads to the parameterization (\ref{ParamSO4}) which ensures that the metric (\ref{MetrSO4a}) has four cyclic directions corresponding to the Cartan subgroup of $SO(4)\times SO(4)$, and this is the maximal number of the cyclic directions for the $SO(4)$ WZW model.

Substituting the parameterization (\ref{ParamSO4}) into (\ref{MetrSO4a}), we arrive at the metric
\bea\label{MetrSO4}
ds^2&=&\frac{4k}{\pi}\left[\frac{dX_1^2}{(X_1^2+1)^2}+\frac{dX_2^2}{(X_2^2+1)^2}\right]+
\frac{k}{\pi}\left[d\alpha_L^2+d\beta_L^2+d\alpha_R^2+d\beta_R^2\right]\\
&&+\frac{k}{\pi}\left[\frac{8 X_1X_2(d\alpha_L d\beta_R+d\alpha_R d\beta_L)}{(1+X_1^2)(1+X_2^2)}+
\frac{2(1-X_1^2)(1-X_2^2)}{(1+X_1^2)(1+X_2^2)}(d\alpha_L d\alpha_R+d\beta_L d\beta_R)
\right].\nonumber
\eea
The Kalb--Ramond field is given by 
\bea\label{BfieldSO4}
B&=&\frac{k}{\pi}\frac{(X_1-X_2)^2}{(X_1^2+1)(X_2^2+1)}(d\alpha_L+d\beta_L)\wedge (d\alpha_R+d\beta_R)\nn
&&\frac{k}{\pi}\frac{(X_1+X_2)^2}{(X_1^2+1)(X_2^2+1)}(d\alpha_L-d\beta_L)\wedge (d\alpha_R-d\beta_R).
\eea
As expected, this geometry has four cyclic coordinates $(\alpha_L,\beta_L,\alpha_R,\beta_R)$, so solutions of the Helmholtz equation
\bea\label{Helm}
\nabla^2\Phi=-\Lambda\Phi
\eea
can be written in the form
\bea
\Phi=e^{in_1\alpha_L+in_2\beta_L+in_3\alpha_R+in_4\beta_R}\,{\tilde\Phi}(X_1,X_2).
\eea
A direct inspection of the metric (\ref{MetrSO4}) and its inverse shows that variables $(X_1,X_2)$ do not separate in the equation (\ref{Helm}). On the other hand, since $SO(4)=SU(2)_1\times SU(2)_2$, there is an alternative parameterization of the group element where the full separation is guaranteed. Specifically, writing an element of $SU(2)_1$ as
\bea
g_1= \begin{bmatrix}
	e^{i\gamma_1}&0\\
	0& e^{-i\gamma_1}
\end{bmatrix}\begin{bmatrix}
	\cos\mu_1&\sin\mu_1\\
	-\sin\mu_1&\cos\mu_1
\end{bmatrix}\begin{bmatrix}
	e^{i\tau_1}&0\\
	0& e^{-i\tau_1}
\end{bmatrix}
\eea
and using a similar expression for $g_2$, we find 
\bea
ds^2&=&\frac{k}{\pi}\sum_{j=1}^2\left[d\mu_j^2+(d\gamma_j+d\tau_j)^2-4\sin^2\mu_j d\gamma_jd\tau_j\right]\\
B&=&\frac{k}{\pi}\sum \cos(2\mu_j)d\gamma_j\wedge d\tau_j\nonumber
\eea
Comparison of the $B$--field with (\ref{BfieldSO4}) suggests the system of separable coordinates:
\bea\label{YcoordGroup}
y_1=1-\frac{2(X_1+X_2)^2}{(X_1^2+1)(X_2^2+1)},\quad 
y_2=1-\frac{2(X_1-X_2)^2}{(X_1^2+1)(X_2^2+1)}
\eea
In the coordinates (\ref{YcoordGroup}) the geometry (\ref{MetrSO4})--(\ref{BfieldSO4}) becomes 
\bea\label{MetrSO4y}
ds^2&=&\frac{k}{2\pi}\left[\frac{dy_1^2}{1-y_1^2}+
(d\alpha_L-d\beta_L)^2+(d\alpha_R-d\beta_R)^2+
2y_1(d\alpha_L-d\beta_L)(d\alpha_R-d\beta_R)\right]\nn
&+&\frac{k}{2\pi}\left[\frac{dy_2^2}{1-y_2^2}+
(d\alpha_L+d\beta_L)^2+(d\alpha_R+d\beta_R)^2+
2y_2(d\alpha_L+d\beta_L)(d\alpha_R+d\beta_R)\right]\nonumber\\
\\
B&=&\frac{k}{2\pi}\left[(1-y_2)(d\alpha_L+d\beta_L)\wedge (d\alpha_R+d\beta_R)+(1-y_1)(d\alpha_L+d\beta_L)\wedge (d\alpha_R+d\beta_R)\right]\nonumber
\eea
Imposing a separable ansatz for the scalar field,
\bea\label{PsiSO4a}
\Phi=e^{in_1(\alpha_L-\beta_L)+in_2(\alpha_R-\beta_R)+i{\tilde n}_1(\alpha_L+\beta_L)+
i{\tilde n}_2(\alpha_R+\beta_R)}B_1(y_1){B}_2(y_2),
\eea
we arrive at the system of ODEs governing functions $(B_1,B_2)$:
\bea\label{SO4scalODE}
&&\frac{d}{dy_1}\left[(y_1^2-1)B_1'\right]+\left[-\nu_1+\frac{(n_1+n_2)^2}{2(1+y_1)}
+\frac{(n_1-n_2)^2}{2(1-y_1)}\right] {B_1}=0\nn
\\
&&\frac{d}{dy_2}\left[(y_2^2-1){B}_2'\right]+\left[-\nu_2+\frac{({\tilde n}_1+{\tilde n}_2)^2}{2(1+y_2)}
+\frac{({\tilde n}_1-{\tilde n}_2)^2}{2(1-y_2)}\right] {B}_2=0.\nonumber
\eea
The eigenvalue of the Helmholtz equation (\ref{Helm}) is given by 
\bea\label{tempLabelB}
\Lambda=\frac{2\pi}{k}\left[\nu_1+\nu_2\right].
\eea
The normalizable solution of the first equation in (\ref{SO4scalODE}) is
\bea\label{B1hyper}
B_1(y_1)=(1+y_1)^{\frac{n_+}{2}}(1-y_1)^{\frac{n_-}{2}}F\left[-p,1+p+n_++n_-;1+n_-;\frac{1-y_1}{2}\right],
\eea
where $p$ is a non--negative integer and 
\bea
n_\pm =|n_1\pm n_2|,\quad \nu_1=\frac{1}{4}(2p+1+n_++n_-)^2-\frac{1}{4}\,.
\eea
Rewriting the last relation in a suggestive form
\bea\label{nu1SO4}
\nu_1=j_1(j_1+1),\quad j_1=p+\frac{n_++n_-}{2},
\eea
we conclude that the eigenvalue $\nu_1$ is equal to the Casimir parameter for a representation of $SU(2)$ described by a Young tableau with $2j$ boxes. There is a similar expression for 
$\nu_2$, 
\bea\label{nu2SO4}
\nu_2=j_2(j_2+1),\quad j_2={\tilde p}+\frac{{\tilde n}_++{\tilde n}_-}{2},
\eea
and single--valuedness of (\ref{PsiSO4a}) as a function of $(\alpha_L,\beta_L,\alpha_R,\beta_R)$
implies that $(j_1,j_2)$ must be either integers or half--integers. The equation (\ref{tempLabelB}) gives the expression for the eigenvalue $\Lambda$ in terms of the Casimir of $SO(4)$
\bea\label{tempLamC2}
\Lambda=\frac{2\pi}{k}C_2(R).
\eea
This agrees with the general expression for the eigenvalues of scalars on the WZW backgrounds  \cite{PS}. In the next subsection will extend these results to the vector field, and sections \ref{SecSO42},  \ref{SecSO422} will focus on extensions of (\ref{PsiSO4a}) and (\ref{tempLamC2}) to various cosets.

\subsection{Vector fields on the $SO(4)$ WZW model}
\label{SecVecSO4}

We have demonstrated separation of variables in the Helmholtz equation, and the next three subsections 
we will show that such separation persists for the analogous equation for the vector field: 
\bea\label{VecHelm1}
e^{2\sigma\phi}\nabla_\mu\left[e^{-2\sigma\phi}{\cal F}^{\mu\nu}\right]+\Lambda A^\nu=0.
\eea
Here following \cite{KubSen}, we introduced a modified field strength in the presence of torsion:
\bea\label{VecHelm2}
{\cal F}_{\mu\nu}=\d_\mu A_\nu-\d_\nu A_\mu+\zeta H_{\mu\nu\sigma}A^\sigma=
F_{\mu\nu}+\zeta H_{\mu\nu\sigma}A^\sigma
\eea
Let us consider equations  (\ref{VecHelm1})--(\ref{VecHelm2}) on $SO(4)$ and its cosets.

\subsubsection{Vector fields on product spaces}
\label{SecSubSub1}

Since $SO(4)=SU(2)\times SU(2)$ and $SO(4)/[SO(2)\times SO(2)]= [SU(2)/U(1)]^2$ correspond to product spaces, we begin with a general discussion on vector fields on such manifolds. Specifically, we consider a geometry that has the form
\bea\label{ProductSpace}
ds^2&=&g_{ij}dx^idx^j+h_{ab}dy^ady^b,\quad e^{-2\phi}=f(x){\tilde f}(y),\\
H&=&\frac{1}{6}H_{ijk}dx^i\wedge dx^j\wedge dx^k+
\frac{1}{6}H_{abc}dy^a\wedge dy^b\wedge dy^c\nonumber
\eea
Equations (\ref{VecHelm1})--(\ref{VecHelm2}) on such a space become
\bea\label{MaxwSyst1}
&&\frac{1}{f^\sigma\sqrt{g}}\d_i\left[f^\sigma\sqrt{g}{\cal F}^{ij}\right]+
\frac{1}{{\tilde f}^\sigma\sqrt{h}}\d_a\left[{\tilde f}^\sigma\sqrt{h}{F}^{aj}\right]
+\Lambda A^j=0\nn
&&\frac{1}{f^\sigma\sqrt{g}}\d_i\left[f^\sigma\sqrt{g}{F}^{ib}\right]+
\frac{1}{{\tilde f}^\sigma\sqrt{h}}\d_a\left[{\tilde f}^\sigma\sqrt{h}{\cal F}^{ab}\right]
+\Lambda A^b=0
\eea
To separate variables between $x$-- and $y$--spaces, we impose the ansatz
\bea\label{VecSeparAnstz}
A={\tilde B}(y) C_i(x) dx^i+{B}(x) C_a(y) dy^a\,.
\eea
There are three types of separable solutions (\ref{VecSeparAnstz}):
\begin{enumerate}[(a)]
\item {\bf Vector fields on the $x$--space:}
\bea\label{VectSeparOne}
A={\tilde B}(y) C_i(x) dx^i
\eea
Substitution into the system (\ref{MaxwSyst1}) gives
\bea
&&\frac{1}{f^\sigma\sqrt{g}}\d_i\left[f^\sigma\sqrt{g}{\cal F}^{ij}\right]+
\left[\frac{1}{{\tilde f}^\sigma\sqrt{h}}\d_a\left[{\tilde f}^\sigma\sqrt{h}{h}^{ab}\d_b{\tilde B}\right]
+\Lambda {\tilde B}\right] C^j=0\nn
&&\frac{1}{f^\sigma\sqrt{g}}\d_i\left[f^\sigma\sqrt{g}g^{ij}{C}_{j}\right]=0\nonumber
\eea
This leads to a system of two eigenvalue problems for decoupled ODEs:
\bea\label{VectBranch1ODE}
\frac{1}{{\tilde f}^\sigma\sqrt{h}}\d_a\left[{\tilde f}^\sigma\sqrt{h}{h}^{ab}\d_b{\tilde B}\right]+
{\tilde \lambda}_{scalar} {\tilde B}=0,\quad
\frac{1}{f^\sigma\sqrt{g}}\d_i\left[f^\sigma\sqrt{g}{\cal C}^{ij}\right]+
\lambda_{vector} C^j=0,
\eea
where ${\cal C}_{ij}\equiv\frac{1}{\tilde B}{\cal F}_{ij}$ is the modified field strength corresponding to the
potential $C_i$:
\bea\label{CtildStren}
{\cal C}_{ij}=\d_i C_j-\d_j C_i+\zeta H_{ijk}C^k\,.
\eea
The system (\ref{VectBranch1ODE}) leads to the eigenvalue 
\bea\label{LamVecOne}
\Lambda={\tilde \lambda}_{scalar}+\lambda_{vector}\,.
\eea
The vector field $C_i$ satisfies a constrant
\bea\label{LorentzOne}
\frac{1}{f^\sigma\sqrt{g}}\d_i\left[f^\sigma\sqrt{g}g^{ij}{C}_{j}\right]=0,
\eea
which ensures that the number of degrees of freedom covered by the ansatz (\ref{VectSeparOne}) is 
$(\mbox{dim}_x-1)$.
\item {\bf Vector fields on the $y$--space:}
\bea\label{VectSeparTwo}
A={B}(x) {\tilde C}_a(y) dy^a\,.
\eea
As before, substitution into the system (\ref{MaxwSyst1}) leads to a two eigenvalue problems for decoupled ODEs\footnote{The field strength ${\cal C}_{ab}$ is defined by the counterpart of (\ref{CtildStren}), 
${\tilde{\cal C}}_{ab}=\d_a {\tilde C}_b-\d_b {\tilde C}_a+\zeta H_{abc}{\tilde C}^c$, and it is related to the relevant components of (\ref{VecHelm2}) by ${\tilde{\cal C}}_{ab}=\frac{1}{B}{\cal F}_{ab}$.},
\bea\label{VectBranch2ODE}
\frac{1}{{f}^\sigma\sqrt{g}}\d_i\left[{f}^\sigma\sqrt{g}{g}^{ij}\d_j{B}\right]+
{\lambda}_{scalar} {B}=0,\quad
\frac{1}{f^\sigma\sqrt{g}}\d_a\left[f^\sigma\sqrt{h}\,{\tilde{\cal C}}^{ab}\right]+
\lambda_{vector} {\tilde C}^b=0,
\eea
and the eigenvalue $\Lambda$ is given by
\bea
\Lambda={\lambda}_{scalar}+{\tilde \lambda}_{vector}\,.
\eea
The vector field ${\tilde C}_a$ satisfies a constraint
\bea\label{LorentzTwo}
\frac{1}{{\tilde f}^\sigma\sqrt{h}}\d_a\left[{\tilde f}^\sigma\sqrt{h}h^{ab}{\tilde C}_{b}\right]=0,
\eea
so the ansatz (\ref{VectSeparOne}) describes  $(\mbox{dim}_y-1)$ degrees of freedom.
\item {\bf The scalar mode:}
\bea\label{VectScalBranch}
A={\tilde B}(y) d C(x)+{B}(x) d{\tilde C}(y).
\eea
Substitution to the system (\ref{MaxwSyst1}) leads to the consistency conditions\footnote{We used the equations of motion for the three--form $H_{\mu\nu\la}$.}
\bea\label{VectScalBranchEqn2}
C(x)=B(x),\quad {\tilde C}(y)=\mu{\tilde B}(y)
\eea
and to system of ODEs:
\bea\label{VectScalBranchODE}
&&\frac{1}{{f}^\sigma\sqrt{g}}\d_i\left[{f}^\sigma\sqrt{g}{g}^{ij}\d_j{B}\right]+
{\lambda}_{scalar} {B}=0,\nn
&&\frac{1}{{\tilde f}^\sigma\sqrt{h}}\d_a\left[{\tilde f}^\sigma\sqrt{h}{h}^{ab}\d_b{\tilde B}\right]+
{\tilde \lambda}_{scalar} {\tilde B}=0.
\eea
The eigenvalue $\Lambda$ and the parameter $\mu$ are given by
\bea\label{LamMuVec}
\Lambda={\lambda}_{scalar}+{\tilde \lambda}_{scalar}\,\quad
\mu=-\frac{{\lambda}_{scalar}}{{\tilde \lambda}_{scalar}}
\eea
If $\mu=1$, then the ansatz (\ref{VectScalBranch}) describes a pure gauge: it gives $\Lambda=0$, but 
functions $(B,{\tilde B})$ remain arbitrary, and they are not constrained by the system (\ref{VectScalBranchODE}). 
\end{enumerate}
To summarize, the separable ansatz (\ref{VecSeparAnstz}) describes 
\bea
\mbox{dim}_x+\mbox{dim}_y-1
\eea
modes with non--zero values of $\Lambda$, and the eigenvalues are given by 
\bea
\Lambda=\left\{{{\tilde \lambda}_{scalar}+\lambda_{vector}},{{\lambda}_{scalar}+{\tilde \lambda}_{scalar}},{\lambda}_{scalar}+{\tilde \lambda}_{vector}\right\}\,.
\eea
Therefore, for find the complete spectrum of the equation (\ref{VecHelm1}) on the product space (\ref{ProductSpace}), it is suffient to determine the scalar and vector eigenvalues 
$({\lambda}_{scalar},{\tilde \lambda}_{scalar},\lambda_{vector},{\tilde \lambda}_{vector})$ on the individual blocks. We will now solve this problem for $SO(4)= SU(2)\times SU(2)$, and we will analyze 
$SO(4)/[SO(2)\times SO(2)]=[SU(2)/U(1)]^2$ in section \ref{SecSO422}.

\subsubsection{Vector modes on $SU(2)$}
\label{SecSubSub2}

To evaluate $({\lambda}_{scalar},{\tilde \lambda}_{scalar},\lambda_{vector},{\tilde \lambda}_{vector})$ and the corresponding eigenfunctions for $SO(4)$, we recall the geometry (\ref{MetrSO4y}) in the $(y_1,y_2)$ coordinates. Comparing it to the general product space (\ref{ProductSpace}), we conclude the $f={\tilde f}=1$, and that the scalar equations (\ref{VectScalBranchODE}) reduce to (\ref{SO4scalODE}) with identification
\bea
&&B=e^{in_1(\alpha_L-\beta_L)+in_2(\alpha_R-\beta_R)}B_1(y_1),\quad 
{\tilde B}=e^{i{\tilde n}_1(\alpha_L-\beta_L)+
i{\tilde n}_2(\alpha_R-\beta_R)}{B}_2(y_2),\nn
&&\lambda_{scalar}=\frac{2\pi}{k}\nu_1,\quad {\tilde\lambda}_{scalar}=\frac{2\pi}{k}\nu_2\,.
\eea
Next we consider the vector equation from the system (\ref{VectBranch1ODE})\footnote{To simplify the subsequent formulas, we rescaled the eigenvalue in (\ref{VectBranch1ODE}) as $\lambda_{vector}=\frac{2\pi}{k}\la$. Then $\la$ is analogous to $(\nu_1,\nu_2)$, and it will be equal to a product of integers or half--integers.},
\bea\label{temp1Fij}
\frac{1}{\sqrt{g}}\d_i\left[\sqrt{g}{\cal F}^{ij}\right]+
\frac{2\pi}{k}\lambda\, C^j=0,\quad 
{\cal F}_{ij}=\d_i C_j-\d_j C_i+\zeta H_{ijk}C^k\,,
\eea
on the relevant part of the geometry (\ref{MetrSO4y}) 
\bea\label{MetrSO4y1}
ds^2&=&\frac{k}{2\pi}\left[\frac{dy_1^2}{1-y_1^2}+
(d\gamma_L)^2+(d\gamma_R)^2+
2y_1 d\gamma_L d\gamma_R\right],\nn
B&=&\frac{k}{2\pi}(1-y_1) d\gamma_L \wedge  d\gamma_R,\quad 
\gamma_L=\alpha_L-\beta_L,\quad
\gamma_R=\alpha_R-\beta_R\,.
\eea
The full analysis of the equation (\ref{temp1Fij}) is presented in the Appendix \ref{AppSU2}, and here we just outline the logic and write the final result. 
\begin{enumerate}[(i)]
\item The most general separable solution for a vector field in the geometry (\ref{MetrSO4y1}) is given by
\bea\label{CfieldFormSU2}
C_i dx_i=e^{i n_1\gamma_L+in_2\gamma_R}\left[V_y dy_1+V_1 d\gamma_L+V_2 d\gamma_R\right],
\eea
where $(V_y,V_1,V_2)$ are functions of $y_1$, which are mixed in equations  (\ref{temp1Fij}). We are looking for combinations of these components that satisfy decoupled equations, and to get insights into the structure of such combinations, we begin with studying the $\zeta=0$ case.

As demonstrated in the Appendix \ref{AppSU2}, the most general separable solutions of equations (\ref{temp1Fij}) with $\zeta=0$ are given by 
\bea\label{CfieldZeta0v1m}
C_i dx_i=e^{i n_1\gamma_L+in_2\gamma_R}\left[V_y dy_1+
q_+({\hat V}_++{\hat V}_-)d\gamma_L+q_-({\hat V}_+-{\hat V}_-)d\gamma_R\right],
\eea
where
\bea\label{qPMratio}
q_\pm=1\pm\frac{n_1^2-n_2^2}{[\sqrt{\la-n_1^2}+\sqrt{\la-n_2^2}]^2},\quad 
\frac{q_-}{q_+}=\left[\frac{\la-n_2^2}{\la-n_1^2}\right]^{1/2}.
\eea
Equations for the functions $({\hat V}_+,{\hat V}_-)$ decouple, and they are formulated as a system of eigenvalue problems 
\bea\label{VectODEzeta0m}
&&\frac{d}{dy_1}\left[\frac{(1-y_1^2){\hat V}_{+}'}{\la y_1-n_1n_2+\mu}\right]+\frac{\la y_1-n_1n_2-\mu}{\la(y_1^2-1)}{\hat V}_{+}=0\nn
~\\
&&\frac{d}{dy_1}\left[\frac{(1-y_1^2){\hat V}_{-}'}{\la y_1-n_1n_2-\mu}\right]+
\frac{\la y_1-n_1n_2+\mu}{\la(y_1^2-1)}{\hat V}_{-}=0\nonumber
\eea
Here we defined $\mu$ as a conveneint combination of constants $(\la,n_1,n_2)$:
\bea
\mu=\sqrt{(\la-n_1^2)(\la-n_2^2)}\,.
\eea
Note that, even though the modes ${\hat V}_+$ and ${\hat V}_-$ decouple, function
\bea
W_+=\frac{(1-y_1^2){\hat V}_{+}'}{\la y_1-n_1n_2+\mu}
\eea
satisfies the same differential equation as ${\hat V}_-$. Similarly, a function $W_-$ constructed from a derivative of ${\hat V}_-$ satisfies the same equation as ${\hat V}_+$.

\item Interestingly, functions $({\hat V}_+,{\hat V}_-)$ can be expressed in terms of the solutions of 
the scalar equation (\ref{SO4scalODE}). As demonstrated in the Appendix \ref{AppSU2}, any solution of equations (\ref{VectODEzeta0m}) can be written as 
\bea\label{VpmAsB1m}
{\hat V}_\pm(y_1)=C_\pm\left[(1-y_1^2)B_1'+\frac{1}{M}
\left[\la y_1-n_1n_2\pm\mu\right]B_1\right]\,,
\eea
where function $B_1$ satisfies the differential equation (\ref{SO4scalODE}), and parameters $(\la,\nu_1,M)$ are related by
\bea\label{EigenValVecZet0}
\la=M^2,\quad \nu_1=M(M+1)\,,\quad M\ge 0.
\eea
In partucular, this implies that the eigenvalues of the problem (\ref{VectODEzeta0}) are given by $\la=M^2$ with an integer $M$ which is subject to the constraint
\bea
M\ge |n_1+n_2|+|n_1-n_2|\,.
\eea
The last remaining component of the vectror field, $V_y$ is given by 
\bea\label{VyAsB1}
V_y&=&iC_+\frac{n_1 q_-+n_2q_+}{M}\left[B_1'-\frac{n_1n_2-M^2 y_1+\mu}{M(1-y_1^2)}B_1\right]\nn
&+&iC_-\frac{n_2 q_+-n_1q_-}{M}\left[B_1'-\frac{n_1n_2-M^2 y_1-\mu}{M(1-y_1^2)}B_1\right]
\eea
\item
To extend the solution (\ref{CfieldZeta0v1m}), (\ref{VpmAsB1m}), (\ref{EigenValVecZet0}), (\ref{VyAsB1}) to arbitrary values of $\zeta$, we observe that two linear combinations of (\ref{VpmAsB1m}) are especially simple: the ones with 
\bea
C_-=C_+ \quad\mbox{and}\quad C_-=-C_+\,.
\eea
Let us begin with analyzing the first combination by setting $C_-=C_+=\frac{1}{2}$:
\bea\label{VecCplus}
&&{\hat V}_+-{\hat V}_-=\frac{\mu}{M}B_1,\quad
{\hat V}_++{\hat V}_-=\left[(1-y_1^2)B_1'+\frac{1}{M}
\left[\la y_1-n_1n_2\right]B_1\right],\nn
&&V_y=\frac{in_2 q_+}{M}\left[B_1'+\frac{\la y_1-n_1n_2}{M(1-y_1^2)}B_1\right]-
\frac{in_1 q_-}{M^2}\frac{\mu}{(1-y_1^2)}B_1
\eea
Comparing with (\ref{CfieldZeta0v1m}), we observe the $C_{\gamma_R}$ component is given by the scalar wavefunction $B_1$. This suggests that it might be useful to write the vector field $C_i$ in terms of the frames\footnote{Once $e_{\bf 3}^\mu$ is fixed by the observation above, the components of 
$e_{\bf \pm}^\mu$ are uniquely determined up to the overall factors.}
\bea\label{LeftFramesMain}
e_{\bf 3}^\mu\d_\mu=\d_{\gamma_R},\quad e^\mu_\pm\d_\mu
=-\frac{e^{\mp i\gamma_R}}{2\sqrt{1-y_1^2}}\left[(1-y_1^2)\d_{y_1}\pm i(y_1\d_{\gamma_R}-
\d_{\gamma_L})\right]
\eea
Evaluating various projections of the field (\ref{VecCplus}), we find remarkably simple relations:
\bea\label{AnstzForSU2}
e_{\bf 3}^\mu C_\mu=a_3e_{\bf 3}^\mu \d_\mu Z,\quad
e_{\bf +}^\mu C_\mu=a_+e_{\bf +}^\mu \d_\mu Z,\quad
e_{\bf -}^\mu C_\mu=a_-e_{\bf -}^\mu \d_\mu Z
\eea
where\footnote{Recall that $q_-\mu=(\la-n_2^2)q_+$.}
\bea\label{a3aPmCoef}
Z=B_1 e^{i n_1\gamma_L+in_2\gamma_R},\quad a_3=\frac{q_+(\la-n^2_2)}{in_2M},\quad
a_\pm=\frac{i(n_2\mp M)q_+}{M}\,.
\eea
Note that the expressions \eqref{AnstzForSU2} are reminiscent of the ansatz for solving the Maxwell's equations in the Myers-Perry geometry \cite{LMaxw}. 

\item The analysis of the second polarization, $C_+=-C_-=\frac{1}{2}$, can be performed in a similar fashion.  In this case the conveneint frames are 
\bea\label{AnstzForSU2b2}
{\tilde e}_{\bf 3}^\mu\d_\mu=\d_{\gamma_L},\quad {\tilde e}^\mu_\pm\d_\mu
=-\frac{e^{\pm i\gamma_L}}{2\sqrt{1-y_1^2}}\left[(1-y_1^2)\d_{y_1}\mp i(y_1\d_{\gamma_L}-
\d_{\gamma_R})\right],
\eea
and the counterpart of the ansatz (\ref{AnstzForSU2}) with 
$({\tilde e}_{\bf 3}^\mu,{\tilde e}_{+}^\mu,{\tilde e}_{-}^\mu,{\tilde a}_3,{\tilde a}_+,{\tilde a}_-)$ gives 
\bea\label{a3aPmCoefb2}
{\tilde a}_3=\frac{q_-(\la-n^2_1)}{in_1M},\quad
{\tilde a}_\pm=\frac{i(n_1\mp M)q_-}{M}\,,\quad \lambda=M^2.
\eea
It turns out that to extend the results to non--zero values of $\zeta$, it is convenient to choose a different route. Writing the counterpart of the first line in (\ref{VecCplus}) for $C_+=-C_-=\frac{1}{2}$,
\bea\label{VecCplusMin}
{\hat V}_+-{\hat V}_-=\left[(1-y_1^2)B_1'+\frac{1}{M}
\left[\la y_1-n_1n_2\right]B_1\right],\quad {\hat V}_++{\hat V}_-=\frac{\mu}{M}B_1\,.
\eea
we observe that the function
\bea
{\tilde B}_1\equiv -\frac{M}{\mu}\left[(1-y_1^2)B_1'+\frac{1}{M}\left[\la y_1-n_1n_2\right]B_1\right]
\eea
satisfies the scalar equation (\ref{SO4scalODE}) with ${\tilde\nu}_1=M(M-1)$. Furthermore, in terms of ${\tilde B}_1$, relations (\ref{VecCplusMin}) become
\bea
{\hat V}_+-{\hat V}_-=-\frac{\mu}{M}{\tilde B}_1,\quad
 {\hat V}_++{\hat V}_-=\left[(1-y_1^2){\tilde B}_1'-\frac{1}{M}
\left[\la y_1-n_1n_2\right]{\tilde B}_1\right]\,.
\eea
These expressions can be obtained from (\ref{VecCplusMin}) by a formal replacement 
\bea
B_1\rightarrow {\tilde B}_1,\quad M\rightarrow -M,\quad \nu_1\rightarrow{\tilde\nu}_1\,.
\eea
and the same replacement works for $V_y$ as well. Therefore, the $C_-=-C_+$ polarization can still be described by the ansatz (\ref{AnstzForSU2}), but relations (\ref{a3aPmCoef}) should be replaced by
\bea\label{a3aPmCoef2brn}
Z={\tilde B}_1 e^{i n_1\gamma_L+in_2\gamma_R},\quad a_3=\frac{q_+(\la-n^2_2)}{in_2M},\quad
a_\pm=\frac{i(n_2\pm M)q_+}{M}\,,\quad M\ge 0\,.
\eea 
Alternatively, we can keep only expressions (\ref{AnstzForSU2})--(\ref{a3aPmCoef}), but allow parameter 
$M$ to take both positive and negative values\footnote{Recall that $M$ was defined in (\ref{EigenValVecZet0}) as a square root of $\la$.}.
\item
In the case of general $\zeta$, we impose the ansatz (\ref{AnstzForSU2}) with 
\bea\label{ZfuncZeta}
Z=B_1 e^{i n_1\gamma_L+in_2\gamma_R},\quad \nu_1=M(M+1),
\eea
and undetermined constants $(a_3,a_+,a_-)$. In accordance with the discussion from item (iv), parameter 
$M$ can take positive and negative values, so to recover both polarizations, every scalar mode is used twice. 
As demonstrated in the Appendix \ref{AppSU2}, the constants $(a_3,a_+,a_-)$ obey the same relations (\ref{a3aPmCoef}),
\bea\label{AvaluesZeta}
a_\pm=\frac{n_2}{n_2\pm M}a_3\,,
\eea
even for a non-vanishing $\zeta$, but the expression for the eigenvalue $\lambda$ in terms of the parameter $M$  is modified as 
\bea\label{VecEigen}
\lambda=M(M-\zeta)\,.
\eea
In particular, for $\zeta=\pm 1$ the sets of scalar and vector eigenvalues are identical. 
\item Finally, there are solutions with $\la_{vec}=0$, which correspond to a pure gauge:
\bea\label{CpureGauge}
C_i dx^i = df.
\eea
For $\zeta=0$, equations (\ref{temp1Fij}) are trivially satisfied since $F_{ij}=0$. In the case of a nonzero 
$\zeta$, the modified field strength does not vanish, but since $H_{ijk}$ is proportional to the volume form, the field ${\cal F}_{ij}$ is divergence-free:
\bea
{\cal F}_{ij}=\zeta H_{ijk}C^k\quad\Rightarrow\quad
\frac{1}{\sqrt{g}}\d_i\left[\sqrt{g}{\cal F}^{ij}\right]=
\frac{\zeta}{\sqrt{g}}\d_i\left[\sqrt{g}{H}^{ijk}C_k\right]=\zeta {H}^{ijk}\d_iC_k=0.
\eea
Therefore, equations (\ref{temp1Fij}) with $\la=0$ are satisfied by the vector field (\ref{CpureGauge}) with an arbitrary $f$. The ansatz (\ref{AnstzForSU2}) with $a_3=a_+=a_-$ and an arbitrary function $Z$ covers all such solutions.
\end{enumerate}
The construction described here gives {\it the most general} separable solution of equations (\ref{temp1Fij}), and a priori it is not obvious that the Lorentz constraint (\ref{VectSeparOne}) would be satisfied. Remarkably, this constraint follows from the ansatz (\ref{AnstzForSU2}) and equations (\ref{temp1Fij}), without additional assumptions. This implies that the solution ((\ref{ZfuncZeta}),(\ref{AvaluesZeta}),(\ref{VecEigen})) can be used to build the vector modes on the product space $SO(4)= SU(2)\times SU(2)$ using the procedure described in section \ref{SecSubSub1}.

\subsubsection{Summary of the vector fields on $SO(4)$}
\label{SecSubSub3}

Let us now combine the discussion from sections \ref{SecSubSub1} and \ref{SecSubSub2} to describe separation of variables for vector fields on $SO(4)$. We are looking for solutions of the eigenvalue problem (\ref{VecHelm1}) with ${\cal F}$ given by (\ref{VecHelm2}) on the geometry 
(\ref{MetrSO4y}). In this case the $x$ and $y$ coordinates defined in (\ref{ProductSpace}) are given by
\bea
x=\{y_1,\alpha_L-\beta_L,\alpha_R-\beta_R\},\quad 
y=\{y_2,\alpha_L+\beta_L,\alpha_R+\beta_R\}\,.
\eea
According to the general discussion from subsection \ref{SecSubSub1}, there are three types of separable vector modes:
\begin{enumerate}[(a)]
\item {\bf Vector fields on the $x$--space:}\\
The ansatz for the vector field has the form
\bea\label{VectSeparOne1O4}
A={\tilde B}(y) C_i(x) dx^i,
\eea
Then the eigenvalue problem (\ref{VecHelm1}) ensures that the functions $({\tilde B},C_i)$ satisfy the system of differential equations (\ref{VectBranch1ODE}) with the constraint (\ref{LorentzOne}). 

In the case of $SO(4)= SU(2)\times SU(2)$, the scalar function has the form
\bea
{\tilde B}(y)=e^{i{\tilde n}_1(\alpha_L+\beta_L)+i{\tilde n}_2(\alpha_R+\beta_R)}{B}_2(y_2),
\eea
and $B_2(y_2)$ satisfies the second equation in (\ref{SO4scalODE}). 
The components of the vector field $C_i$ have the form (\ref{AnstzForSU2}),
\bea\label{AnstzForSU2a}
e_{\bf 3}^\mu C_\mu=a_3e_{\bf 3}^\mu \d_\mu Z,\quad
e_{\bf +}^\mu C_\mu=a_+e_{\bf +}^\mu \d_\mu Z,\quad
e_{\bf -}^\mu C_\mu=a_-e_{\bf -}^\mu \d_\mu Z,
\eea
with frames (\ref{LeftFramesMain}). Function $Z$ given by 
(\ref{ZfuncZeta})\footnote{Recall that 
$\gamma_L=\alpha_L-\beta_L$, 
$\gamma_R=\alpha_R-\beta_R$.},
\bea
Z=B_1(y_1) e^{i n_1\gamma_L+in_2\gamma_R},
\eea
where $B_1$ is a solution of the first equation in (\ref{SO4scalODE}) with 
$\nu_1=j_1(j_1+1)$.
The eigenvalues of the vector equation (\ref{VecHelm1}) have the form (\ref{LamVecOne}),
\bea\label{LamVecOneV2}
\Lambda={\tilde \lambda}_{scalar}+\lambda_{vector}\,,
\eea
with
\bea\label{LambdaVec}
{\tilde \lambda}_{scalar}=\frac{2\pi}{k}\nu_2=\frac{2\pi}{k}j_2(j_2+1),\quad 
\lambda_{vector}=\frac{2\pi}{k}j_1(j_1-\zeta).
\eea
Note that for $\zeta=\pm 1$ the scalar and vector spectra, (\ref{tempLabelB}) and (\ref{LamVecOneV2}), are identical. 
The ansatz (\ref{VectSeparOne1O4}) describes two physical degrees of freedom.
\item {\bf Vector fields on the $y$--space:}\\
This situation is analogous to the case (a) with a replacement 
\bea
x\rightarrow y,\quad {\tilde \lambda}_{scalar}\rightarrow {\lambda}_{scalar},\quad
{\lambda}_{vector}\rightarrow {\tilde\lambda}_{vector}
\eea  
For example, the ansatz for the vector field is
\bea\label{VectSeparOne2}
A={B}(x) {\tilde C}_a(y) dy^a\,.
\eea
and functions $({B},{\tilde C}_a)$ satisfy the system of differential equations (\ref{VectBranch2ODE}) with the constraint (\ref{LorentzTwo}). The vector ${\tilde C}_a$ and the scalar $B(x)$ are given by 
\bea
&&{\tilde e}_{\bf 3}^a {\tilde C}_a=a_3e_{\bf 3}^a \d_a {\tilde Z},\quad
{\tilde e}_{\bf +}^a C_a=a_+{\tilde e}_{\bf +}^a \d_a {\tilde Z},\quad
{\tilde e}_{\bf -}^a C_a=a_-{\tilde e}_{\bf -}^a \d_a {\tilde Z},\\
&&{B}(x)=e^{i{n}_1(\alpha_L-\beta_L)+i{n}_2(\alpha_R-\beta_R)}{B}_1(y_1),\quad
{\tilde Z}(y)=e^{i{\tilde n}_1(\alpha_L+\beta_L)+i{\tilde n}_2(\alpha_R+\beta_R)}{B}_2(y_2),
\nonumber
\eea
and $(B_1(y_1),B_2(y_2))$ satisfy equations (\ref{SO4scalODE}). The eigenvalues of the equation (\ref{VecHelm1}) are
\bea\label{LamVecOne1}
\Lambda={{\lambda}_{scalar}+\tilde\lambda_{vector}}\,.
\eea
with 
\bea\label{LambdaVecBr2}
{\lambda}_{scalar}=\frac{2\pi}{k}j_1(j_1+1),\quad 
{\tilde\lambda}_{vector}=\frac{2\pi}{k}j_2(j_2-\zeta).
\eea
The ansatz (\ref{VectSeparOne2}) describes two physical degrees of freedom.
\item {\bf The scalar mode:}\\
The ansatz for the gauge field is given by (\ref{VectScalBranch})--(\ref{VectScalBranchEqn2}):
\bea\label{VectScalBranch1}
A={\tilde B}(y) d B(x)+\mu {B}(x) d{\tilde B}(y), 
\eea
and in the $SO(4)$ case, 
\bea
B=e^{in_1(\alpha_L-\beta_L)+in_2(\alpha_R-\beta_R)}B_1(y_1),\quad 
{\tilde B}=e^{i{\tilde n}_1(\alpha_L+\beta_L)+
i{\tilde n}_2(\alpha_R+\beta_R)}{B}_2(y_2)
\eea
Functions $B_1$ and $B_2$ satisfy equations (\ref{SO4scalODE}), the eigenvalue $\Lambda$ and parameter $\mu$ are given by (\ref{LamMuVec})
\bea\label{LamMuVec1}
\Lambda={{\lambda}_{scalar}+{\tilde \lambda}_{scalar}}\,,\quad
\mu=-\frac{{\lambda}_{scalar}}{{\tilde \lambda}_{scalar}}
\eea
As in the general case discussed in section \ref{SecSubSub1}, $\mu=1$ corresponds to a pure gauge, which gives $\Lambda=0$ and arbitrary functions $(B,{\tilde B})$ (the equations
(\ref{SO4scalODE}) are not required).  
\end{enumerate}
To summarize, application of the separable ansatz (\ref{VecSeparAnstz}) to $SO(4)$ describes five physical degrees of freedom, and the set of eigenvalues is given by 
\bea\label{SO4eigenSum}
\Lambda=\left\{{{\tilde \lambda}_{scalar}+\lambda_{vector}},{{\lambda}_{scalar}+{\tilde \lambda}_{scalar}},{\lambda}_{scalar}+{\tilde \lambda}_{vector}\right\}\,.
\eea
The individual ingredients are specified by two numbers $(j_1,j_2)$, which can be either both integers or both half--integers:
\bea\label{ScalVsVecSpec}
&&{\lambda}_{scalar}=\frac{2\pi}{k}j_1(j_1+1),\quad {\tilde \lambda}_{scalar}=\frac{2\pi}{k}j_2(j_2+1)\nn
&&\lambda_{vector}=\frac{2\pi}{k}j_1(j_1-\zeta),\quad
{\tilde\lambda}_{vector}=\frac{2\pi}{k}j_2(j_2-\zeta).
\eea
The eigenvalues (\ref{SO4eigenSum}) have the standard degeneracy associated with $[U(1)]^4$ quantum numbers $(n_1,n_2,{\tilde n}_1,{\tilde n}_2)$, but this degeneracy is enhances if $\zeta=\pm 1$ when all three ingredients of (\ref{LamMuVec1}) have the same 
$(j_1,j_2)$ dependence.

\subsection{Gauged WZW model for the $SO(4)/SO(2)$ coset}
\label{SecSO42}

Let us now gauge some of the symmetries of $SO(4)$ and study various fields on the resulting backgrounds.
In this subsection we go back to the group element (\ref{ParamSO4}) and gauge the $SO(2)$ symmetry that acts as
\bea
g\rightarrow h g h^{-1},\quad \mbox{where}\quad 
h=\begin{bmatrix}
	q(\mu)&0\\
	0& I_{2\times 2}
\end{bmatrix}
\eea 
Here $\mu$ is the gauge parameter. We choose a convenient gauge by setting  
\bea\label{SO4SO2elem}
X=q_2(\gamma)\mbox{diag}(X_1,X_2), \quad \alpha_R=0,\quad \beta_R=0.
\eea
in the product (\ref{ParamSO4}). Then the metric, the dilaton, and the Kalb--Ramond field of the gauged WZW (gWZW) model become\footnote{The general procedure for constructing the gWZW geometries and its application to the specific case (\ref{SO4SO2elem}) are discussed in the Appendix \ref{AppGWZW}.} 
\bea\label{SO4SO2a}
ds^2&=&\frac{k}{\pi}\left[\frac{4dX_1^2}{(1+X_1^2)^2}+\frac{4dX_2^2}{(1+X_2^2)^2}+d\alpha^2+4\frac{X_1X_2d\alpha d\beta}{X_1^2+X_2^2}+\frac{(1+X_1^2X_2^2)d\beta^2}{(X_1^2+X_2^2)}\right]\nn
&+&\frac{k}{\pi}\left[
\frac{4(X_1^2-X_2^2)^2d\alpha d\gamma}{(1+X_1^2)(1+X_2^2)(X_1^2+X_2^2)}+
\frac{4(X_1^2-X_2^2)^2d\gamma^2}{(1+X_1^2)(1+X_2^2)(X_1^2+X_2^2)}\right],\nn
e^{-2\phi}&=&\frac{2(X_1^2+X_2^2)}{(1+X_1^2)(1+X_2^2)},\quad \alpha=\alpha_L,\quad 
\beta=\beta_L\,.
\eea
The geometry also contains a Kalb--Ramond $B$ field, but the expression for it in the $(X_1,X_2)$ coordinates is not very illuminating.  In terms of coordinates $(y_1,y_2)$ introduced in (\ref{YcoordGroup}), the geometry (\ref{SO4SO2a}) becomes 
\bea\label{SO4SO2}
ds^2&=&\frac{k}{\pi}\left[d\alpha^2+\frac{2(y_1-y_2)}{y_1+y_2-2}d\alpha d\beta-\frac{4(y_1-1)(y_2-1)}{y_1+y_2-2}d\alpha d\gamma-\frac{4(y_1-1)(y_2-1)}{y_1+y_2-2}d\gamma^2\right]\nn
&&+\frac{k}{\pi}\left[-\frac{y_1+y_2+2}{y_1+y_2-2}d\beta^2+\frac{dy_1^2}{2-2y_1^2}+\frac{dy_2^2}{2-2y_2^2}\right]\,,\nn
H&=&\frac{k}{\pi}\left\{(d\alpha+2d\gamma)\wedge d\beta \wedge d\left[\frac{y_2-y_1}{y_2+y_1-2}\right]+
d\alpha\wedge d\gamma \wedge d\left[\frac{2(y_1y_2-1)}{y_2+y_1-2}\right]\right\},\nn
e^{-2\phi}&=&2-y_1-y_2.
\eea 
To demonstrate separation of variables, we observe that the frames in the $(\alpha,\beta,\gamma)$ subspace can be chosen to be
\bea\label{FramesSO4SO2}
&&e_1^\mu\d_\mu=\frac{1}{2}\sqrt{\frac{2-2y_1}{1+y_1}}
\left[\d_\alpha-\d_\beta+\frac{y_1}{1-y_1}\d_\gamma\right],\nn
&&e_2^\mu\d_\mu=\frac{1}{2}\sqrt{\frac{2-2y_2}{1+y_2}}
\left[\d_\alpha+\d_\beta+\frac{y_2}{1-y_2}\d_\gamma\right],\\
&&e_3^\mu\d_\mu=\frac{2}{2}(\d_\alpha-\d_\gamma).\nonumber
\eea
This implies that the inverse metric can be written as 
\bea\label{SO4SO2temp}
\frac{\pi}{k}g^{\mu\nu}\d_\mu\d_\nu=\left[2(1-y_1^2)\d_{y_1}^2+e_1^\mu e_1^\nu\d_\mu\d_\nu\right]+
\left[2(1-y_2^2)\d_{y_2}^2+e_2^\mu e_2^\nu\d_\mu\d_\nu\right]+e_3^\mu e_3^\nu\d_\mu\d_\nu
\eea
The first block depend only on $y_1$, the second block depends only on $y_2$, while the third block contains only constant coefficients. Such structure of frames, has also been encountered in Myers--Perry--AdS black holes in odd dimensions, where it guaranteed separation of the Helmholtz and Hamilton--Jacobi equations \cite{Kub1}, as well as equations for the vector field \cite{LMaxw} and higher forms \cite{LHform}. In the present case, the structure (\ref{SO4SO2temp}) guarantees the full separation of variables in the Hamilton--Jacobi equation 
\bea
g^{\mu\nu}\frac{\d S}{\d x^\mu}\frac{\d S}{\d x^\nu}=\la,
\eea
but not in the Helmholtz equation (\ref{Helpm}). The obstacle comes from a non--separable determinant of the metric: 
\bea
\sqrt{g}=\left(\frac{k}{\pi}\right)^{\frac{5}{2}}\frac{1}{2-y_1-y_2}.
\eea
It turns out that a modified Helmholtz equation
\bea\label{Helpm}
e^{2\sigma\phi}\nabla^2(e^{-2\sigma\phi}\Phi)=-\Lambda\Phi
\eea
is still separable if and only if $\sigma=1$. This special value of $\sigma$ has been already encountered in \cite{PS}, where the general formula for scalar eigenvalues $\Lambda$ on all 
$G/H$ gauged WZW models was derived using algebraic methods, which were applicable only to $\sigma=1$.

Substituting a separable ansatz for the scalar field,
\bea 
\Phi=\exp[i n_1 \alpha+i n_2 \beta+i n_3 (\alpha+2\gamma)] g_1(y_1)g_2(y_2),
\eea 
into the Helmholtz equation (\ref{Helpm}) in the geometry (\ref{SO4SO2a}), and setting $\sigma=1$, we arrive at a system of ODEs: 
\bea\label{SO4SO2ODE}
&&\hskip -1.5cm
\frac{d}{dy_1}\left[(y_1^2-1)\frac{dg_1}{dy_1}\right]+\frac{1}{2}\left[\frac{n_3^2}{1-y_1}+
\frac{(n_1-n_2)^2}{y_1+1}\right]g_1+
\frac{1-\la_1^2}{4}g_1=0,\nn
&&\hskip -1.5cm
\frac{d}{dy_2}\left[(y_2^2-1)\frac{dg_2}{dy_2}\right]+\frac{1}{2}\left[\frac{n_3^2}{1-y_2}+
\frac{(n_1+n_2)^2}{y_2+1}\right]g_2+
\frac{1-\la_2^2}{4}g_2=0.
\eea
The eigenvalue $\Lambda$ is given by 
\bea
{\Lambda}=\frac{2\pi}{k}(\frac{\la_1^2}{4}+\frac{\la_2^2}{4}-\frac{1}{2}-\frac{n_2^2}{2})\,.
\eea
Equations (\ref{SO4SO2ODE}) can be solved in terms of the hypergeometric function, and the result reads 
\bea\label{newEqn42scal}
g_1&=&
(1-y_1)^{\frac{n_3}{2}}(1+y_1)^{\frac{n_-}{2}}F[-k_1,1+n_3+n_-+k_1;1+n_3;\frac{1-y_1}{2}],\nn
g_2&=&
(1-y_2)^{\frac{n_3}{2}}(1+y_2)^{\frac{n_+}{2}}F[-k_2,1+n_3+n_++k_2;1+n_3;\frac{1-y_2}{2}],\nn
{\Lambda}&=&\frac{2\pi}{k}\left[\frac{(2k_1+n_3+n_-+1)^2}{4}+\frac{(2k_2+n_3+
n_++1)^2}{4}-\frac{1+n_2^2}{2}\right].
\eea
Here we assumed that $n_3\ge 0$ and introduced two more non--negative parameters:
\bea
n_+=|n_1+n_2|,\quad n_-=|n_1-n_2|. \nonumber
\eea
It is instructive to compare the expression for the eigenvalue from (\ref{newEqn42scal}) with the  general formula for the gauged WZW models on the $G/H$ cosets \cite{PS}. As demonstrated in \cite{PS}, the eigenfunctions of the scalar field (\ref{Helpm}) on such cosets are specified by a representations of the group $G$ and the subgroup $H$, and the eigenvalues are expressed in terms of the quadratic Casimirs of such representations\footnote{The expression in \cite{PS} is slightly more general, but it reduces to (\ref{tempLamPS}) in the geometric limit which we are discussing here. Also, the algebraic construction of \cite{PS} applies only to the equation (\ref{Helpm}) with $\sigma=1$, and it is remarkable that this equation separates and results in (\ref{newEqn42scal}) precisely for this value of $\sigma$.}:
\bea\label{tempLamPS}
\Lambda=\frac{2\pi}{k}C_2(R_G)-\frac{2\pi}{k}C_2(R_H).
\eea
To recover this formula, we rewrite the expression for $\Lambda$ from (\ref{newEqn42scal}) in a suggestive form:
\bea
&&\Lambda=\frac{2\pi}{k}\left[j_1(j_1+1)+j_2(j_2+1)\right]-\frac{2\pi}{k}\frac{n_2^2}{2},\\
&&j_1=k_1+\frac{n_3+n_-}{2},\quad j_2=k_2+\frac{n_3+n_+}{2}\,.\nonumber
\eea
Note that in  (\ref{nu1SO4}) and (\ref{nu2SO4}) we have already encountered the counterparts of the parameters $(j_1,j_2)$ in the $SO(4)$ case. As expected the general formula (\ref{tempLamPS}) is reproduced.

\subsection{Vector fields on the $SO(4)/SO(2)$ gauged WZW model} 
\label{SecVecO42}

To separate variables in the vector equation (\ref{VecHelm1})
\bea\label{VecEqnSO42}
e^{2\phi}\nabla_\mu\left[e^{-2\phi}{\cal F}^{\mu\nu}\right]+4\Lambda A^\nu=0,\quad 
{\cal F}_{\mu\nu}=\d_\mu A_\nu-\d_\nu A_\mu+\zeta H_{\mu\nu\sigma}A^\sigma\,,
\eea
we implement the idea that has been used to solve Maxwell's equations in the Myers-Perry geometry \cite{LMaxw}. We begin with introducing complex combinations of frames (\ref{FramesSO4SO2}) that depend only on $y_1$ or $y_2$, as well as the constant frame $e_0$:
\bea
e_{1\pm}=e_1\pm i {\tilde e}_1,\quad e_{2\pm}=e_2\pm i {\tilde e}_2,\quad e_0,\quad
{\tilde e}_{1,2}=\sqrt{2(1-y_{1,2}^2)}\d_{y_{1,2}}\,.
\eea
Then we impose an ansatz inspired by our discussion in section \ref{SecVecSO4} and by separation of vector equations in background of the Myers--Perry black holes \cite{LMaxw}
\bea\label{VectAnstz}
&&e_{1\pm}^\mu A_\mu=b_{1\pm}(y_1) e_{1\pm}^\mu\d_\mu {\tilde Z},\quad 
e_{2\pm}^\mu A_\mu=b_{2\pm}(y_2) e_{2\pm}^\mu\d_\mu {\tilde Z},\quad 
e_{0}^\mu A_\mu=b_0 e_{0}^\mu\d_\mu {\tilde Z},\nn
&&{\tilde Z}= e^{in_1\alpha+in_2\beta+i n_3\gamma}{Z}(y_1,y_2).
\eea
Direct substitution into  (\ref{VecEqnSO42}) shows that
the vector equations become separable only for $\zeta=1$, and the results are:
\begin{enumerate}
\item Coefficients $b_{1\pm}$ and $b_{2\pm}$ must be constant. This is consistent with constant eigenvalues of the Killing--Yano tensor corresponding to the metric (\ref{FramesSO4SO2}),
\bea\label{KYTso42}
Y=(e^2\wedge {\tilde e}^2-e^1\wedge {\tilde e}^1)\wedge e^0\,,
\eea
which satisfies equations with twisted connections:
\bea\label{TwistKY1}
&&\nabla^+_{n}Y_{mpq}+\nabla^+_{m}Y_{npq}=0,\quad \Gamma^{d+}_{np}=\Gamma^{d}_{np}+\frac{1}{2}H^d_{np}, \\
&&\nabla^+_n Y_{mpq}=\p_n Y_{mpq}-\Gamma^{d+}_{nm}Y_{dpq}-\Gamma^{d+}_{np}Y_{mdq}-\Gamma^{d+}_{nq}Y_{mpd}=0.\nonumber
\eea

Recall that in the case of the Myers--Perry geometry the metric and the Killing--Yano tensors had the form
\bea
ds^2&=&\left[-l^+_\mu l^-_\nu+\sum_i m^{(i)+}_\mu m^{(i)-}_\nu+n_\mu n_\nu\right] dx^\mu dx^\nu\nn
Y^{(2n-k)}&=&\star[\wedge h^k],\quad h=r l_+\wedge l_-+
\sum_i x_i m^{(i)}_+\wedge m^{(i)}_-\nonumber
\eea
and the counterpart of the ansatz (\ref{VectAnstz}) was\footnote{See \cite{LMaxw} for details and derivation.} 
\bea
&&\hskip -1.5cm l^\mu_\pm A_\mu=\pm\frac{1}{r\pm i\mu}{\hat l}_\pm \Psi,\quad
[m_\pm^{(j)}]^\mu A_\mu=\mp\frac{i}{x_j\pm \mu}{\hat m}^{(j)}_\pm \Psi\,\quad
n^\mu A_\mu=\la \Psi.\nonumber
\eea
In particular, the prefactors in the last equation involved some combinations of the eigenvalues of $h$, and since the eigenvalues of (\ref{KYTso42}) do not depend on $(y_1,y_2)$, the constant values of the coefficients $b_{1\pm}$ and $b_{2\pm}$ in (\ref{VectAnstz}) are not surprising. 
\item Separable function $Z$ obeys a system of ODEs
\bea
\hskip -1cm
&&
\hskip -1cm \frac{1}{1-y_1^2}\frac{\d}{\d y_1}\left[(1-y_1^2)\frac{\d Z}{\d y_1}\right]+\frac{\la_1 Z}{1-y_1^2}-\frac{n_3^2 Z}{4(1-y_1^2)^2}-
\frac{(n_1-n_2)(n_1-n_2-n_3) Z}{2(1-y_1)(1-y_1^2)^2}=0\nn
\\
\hskip -1cm
&&
\hskip -1cm \frac{1}{1-y_2^2}\frac{\d}{\d y_2}\left[(1-y_2^2)\frac{\d Z}{\d y_2}\right]+\frac{\la_2 Z}{1-y_2^2}-
\frac{n_3^2 Z}{4(1-y_2^2)^2}-
\frac{(n_1+n_2)(n_1+n_2-n_3) Z}{2(1-y_2)(1-y_2^2)^2}=0\nonumber
\eea
\item The eigenvalues $\Lambda$ of (\ref{VecEqnSO42}) are given by
\bea\label{SO42verEigen}
\Lambda=\frac{2\pi}{k}\left({\la_1+\la_2}-\frac{n_2^2}{2}\right)\,.
\eea
\item Five coefficients $(b_{1\pm},b_{2\pm},b_0)$ obey one constraint:
\bea
&&(b_{1+}+b_{1-})[4\la_1+1]+(b_{2+}+b_{2-})[4\la_2+1]+4b_0(n_1-n_3)^2-\nn
&&\quad-\sum_\pm b_{1\pm}(n_2+n_3-n_1\pm 1)^2-\sum_\pm b_{2\pm}(n_3-n_1-n_2\pm 1)^2=0
\eea
This constraint follows from the equations (\ref{VecEqnSO42}), and it also ensures the Lorenz condition 
\bea
\nabla_\mu[e^{-2\phi}A^\mu]=0.
\eea 
\end{enumerate}
The metric (\ref{FramesSO4SO2}) also admits another twisted Killing-Yano tensor:
\bea \label{KY2}
\hat{Y}=(\hat{e}^2\wedge {{\tilde e}}^2-\hat{e}^1\wedge {{\tilde e}}^1)\wedge \hat{e}^0\,,
\eea 
where new frames $\hat{e}^2$, $\hat{e}^0$ and $\hat{e}^1$ are 
\bea \label{RightPro}
&&\hat{e}^1=\frac{\sqrt{2-2y_1^2}}{2-y_1-y_2}(d\beta+(1-y_2)d\gamma),\quad \hat{e}^2=-\frac{\sqrt{2-2y_2^2}}{2-y_1-y_2}(d\beta+(1-y_1)d\gamma),\nn
&&\hat{e}^0=d\alpha-\frac{y_1-y_2}{2-y_1-y_2}d\beta+\frac{2(1-y_1)(1-y_2)}{2-y_1-y_2}d\gamma\,.
\eea 
In contrast to \eqref{KYTso42} this twisted Killing-Yano tensor satisfies a different twisted Killing-Yano equation
\bea\label{TwistKY2}
&&\nabla^-_{n}\hat{Y}_{mpq}+\nabla^-_{m}\hat{Y}_{npq}=0,\quad \Gamma^{d-}_{np}=\Gamma^{d}_{np}-\frac{1}{2}H^d_{np}, \\
&&\nabla^-_n \hat{Y}_{mpq}=\p_n \hat{Y}_{mpq}-\Gamma^{d-}_{nm}\hat{Y}_{dpq}-\Gamma^{d-}_{np}\hat{Y}_{mdq}-\Gamma^{d-}_{nq}\hat{Y}_{mpd}=0.\nonumber
\eea 
Therefore another separable ansatz of the vector field equation is possible by replacing $e^1$, $e^2$ and $e^0$ in \eqref{VectAnstz} by $\hat{e}^1$, $\hat{e}^2$ and $\hat{e}^0$, respectively. A direct substitution of this alternative ansatz into \eqref{VecEqnSO42} shows the vector equation becomes separable when $\zeta=-1$. In the Appendix \ref{AppGWZW} we will show these two possible separable ansatze are related to the left and right frames of the gWZW model.

\bigskip

To summarize, in this subsection we have demonstrated separability of the twisted vector equation (\ref{VecEqnSO42}) for two values of the twisting parameter: $\zeta=\pm 1$. In both cases the components of the vector field are given by (\ref{VectAnstz}), but the frames used in these relations are different: $\zeta=1$ corresponds to the left--invariant forms, and $\zeta=-1$ corresponds to the right--invariant ones. The separation of the vector equation is not possible for any other values of $\zeta$, in particular, the standard equation corresponding to $\zeta=0$ does not separate. In the cases when separation is possible, the eigenvalues (\ref{SO42verEigen}) are equal to their scalar counterparts (\ref{newEqn42scal}), so the group theoretic formula (\ref{tempLamPS}) which has been derived for the scalar spectrum, seems to be applicable to vectors with $\zeta=\pm 1$ as well. We have already encountered this phenomenon in section \ref{SecSubSub2}, where the scalar and vector spectra (\ref{ScalVsVecSpec}) agreed precisely for $\zeta=\pm 1$.

\subsection{Scalars and vectors on the $\frac{SO(4)}{SO(2)\times SO(2)}$ gauged WZW model}
\label{SecSO422}
Let us now gauge one more $U(1)$ isometry and study various fields on the resulting $SO(4)/[SO(2)\times SO(2)]$ coset. To do so, we go back to the group element (\ref{ParamSO4}) and gauge the $SO(2)\times SO(2)$ subgroup that acts as
\bea
g\rightarrow h g h^{-1},\quad \mbox{where}\quad 
h=\begin{bmatrix}
	q_2(\mu)&0\\
	0& q_2(\nu)
\end{bmatrix}
\eea 
This leads to the shifts 
\bea
\alpha_{L,R}\rightarrow \alpha_{L,R}\pm \mu,\qquad \beta_{L,R}\rightarrow \beta_{L,R}\pm \nu\,
\eea
in the parameters of (\ref{ParamSO4}), and the gauge can be fixed by setting $\alpha_R=\beta_R=0$. The resulting coset element has the form 
\bea\label{ParamSO4SO22}
g=\begin{bmatrix}
	q_2(\alpha)&0\\
	0&{q}_2(\beta)
\end{bmatrix} \begin{bmatrix}
	I-\frac{2}{1+XX^T}XX^T&\frac{2}{1+XX^T}{X}\\
	-X^T\frac{2}{1+XX^T}& I-\frac{2}{1+X^TX}X^T X
\end{bmatrix},\quad X=\mbox{diag}(X_1,X_2).
\eea
Using the general procedure for constructing the metric of the gauged WZW model\footnote{See Appendix \ref{AppGWZW} for the details.}, we arrive at the geometry
\bea\label{tempSO422geom}
ds^2&=&\frac{k}{\pi} \left[ \frac{4dX_1^2}{(1+X_1^2)^2}+\frac{4dX_2^2}{(1+X_2)^2}+\frac{4X_1X_2 (1+X_1^2)(1+X_2^2)}{(X_1^2-X_2^2)^2}d\alpha d\beta    \right]\nn
&&\qquad +\frac{k}{\pi} \left[\frac{X_2^2+X_1^2(1+4X_2^2+X_1^2X_2^2+X_2^4)}{(X_1^2-X_2^2)^2} \left(   d\alpha^2+d\beta^2\right)\right],\nn
e^{-2\phi}&=&\frac{4(X_1^2-X_2^2)^2}{(1+X_1^2)^2(1+X_2^2)^2}.
\eea
In contrast to the $SO(4)/SO(2)$ coset, the geometry (\ref{tempSO422geom}) does not contain a $B$ field. A sequence of invertible maps,
\bea
X_i=\frac{i(1-w_i)}{1+w_i},\quad w_1=\sqrt{x_1 x_2},\quad w_2=\sqrt{\frac{x_2}{x_1}},\quad
y_i=\frac{1+x_i^2}{2x_i}
\eea
leads to a separable form of the metric : 
\bea\label{SU2tmSU2}
ds^2
&=&\frac{k}{2\pi}\left(\frac{1+y_1}{1-y_1}[d\alpha-d\beta]^2+\frac{1+y_2}{1-y_2}[d\alpha+d\beta]^2+\frac{dy_1^2}{1-y_1^2}+\frac{dy_2^2}{1-y_2^2}\right).\nn
e^{-2\phi}&=&(1-y_1)(1-y_2),\quad \sqrt{g}=\left(\frac{k}{2\pi}\right)^2\frac{1}{(1-y_1)(1-y_2)}
\eea
In contrast to the situations discussed in section \ref{SecSO42}, the scalar equation (\ref{Helpm}) separates for all values of $\sigma$. This is not surprising since the geometry (\ref{SU2tmSU2}) describes two copies of $SU(2)/U(1)$:
\bea\label{SO422ProdStruct}
\frac{SO(4)}{SO(2)\times SO(2)}=\frac{SU(2)_L\times SU(2)_R}{U(1)_L\times U(1)_R}=
\frac{SU(2)_L}{U(1)_L}\times \frac{SU(2)_R}{U(1)_R}
\eea

Imposing a separable ansatz
\bea\label{ScalO422tmp}
\Phi=e^{in_1(\alpha-\beta)+in_2(\alpha+\beta)}Y_1(y_1)Y_2(y_2),
\eea
and substituting the result into (\ref{Helpm}), we arrive at a system of two ODEs:
\bea\label{SO422ODE}
&&\frac{1}{(1-y_1)^{\sigma}}
\frac{d}{dy_1}\left[(1-y_1)^{\sigma}(1+y_1) \frac{dY_1}{dy_1}\right]-
\frac{n_1^2}{1+y_1}Y_1+\frac{\lambda_1}{1-y_1} Y_1=0\nn
\\
&&\frac{1}{(1-y_2)^{\sigma}}
\frac{d}{dy_2}\left[(1-y_2)^{\sigma}(1+y_2) \frac{dY_2}{dy_2}\right]-
\frac{n_2^2}{1+y_2}Y_2+\frac{\lambda_2}{1-y_2} Y_2=0\nonumber
\eea
The eigenvalues of the full problem (\ref{Helpm}) are $\Lambda=\frac{2\pi}{k}(\lambda_1+\lambda_2)$. 

Equations (\ref{SO422ODE}) can be solved in terms of the hypergeometric functions:
\bea
Y_1&=&(1+y_1)^{-n_1}F[-k_1,k_1-2n_1+\sigma;\sigma;\frac{1-y_1}{2}],\nn
Y_2&=&(1+y_2)^{-n_2}F[-k_2,k_2-2n_2+\sigma;\sigma;\frac{1-y_2}{2}].
\eea
The eigenvalues $\Lambda$ in equation (\ref{Helpm}) are given by
\bea\label{LambSO422}
\Lambda=\frac{2\pi}{k}(\lambda_1+\lambda_2),\quad 
\lambda_{1,2}=(k_{1,2}-n_{1,2}+\frac{\sigma}{2})^2-{n_{1,2}^2}
-\frac{\sigma^2}{4}
\eea
Regularity requires $k_1$ and $k_2$ to be non--negative integers. Setting $\sigma=1$ and introducing $j_{1,2}=k_{1,2}-n_{1,2}$ the eigenvalues can be written as
\bea 
\lambda_{1,2}=j_{1,2}(j_{1,2}+1)-n_{1,2}^2.
\eea 
This leads to the expressions for $\Lambda_\pm$ which are consistent with an application of the general formula (\ref{tempLamPS}) for a coset \cite{PS} to the $SU(2)/U(1)$ case. Equation (\ref{LambSO422}) also hints at a potential generalization of the formula (\ref{tempLamPS}) to arbitrary values of $\sigma$. Such generalization indeed exists for all groups and cosets, and it will be discussed elsewhere \cite{LTianNew}. 

\bigskip

The eigenfunctions of the vector field (\ref{VecHelm1}), (\ref{VecHelm2}) follow the pattern outlined in section \ref{SecSubSub1}. In the present case there is no $H$--field, so one does not have to consider $\zeta$--modified vector equations, and the analysis becomes simpler than the one presented in sections \ref{SecSubSub2}, \ref{SecSubSub3}. 

Division of space (\ref{SU2tmSU2}) into two blocks,
\bea
x=\{y_1,\alpha-\beta\},\quad 
y=\{y_2,\alpha+\beta\}\,.
\eea
and application of the general pattern presented in section \ref{SecSubSub1} leads to three types of vector modes:
\begin{enumerate}[(a)]
\item {\bf Vector fields on the $x$--space:}\\
The ansatz for the vector field has the form
\bea\label{VectSeparOne1}
A={\tilde B}(y) C_i(x) dx^i,
\eea
and the eigenvalue problem (\ref{VecHelm1}) leads to the (\ref{VectBranch1ODE}) for the functions 
$({\tilde B},C_i)$. Field $C_i$ must satisfy the constraint (\ref{LorentzOne}) as well, but as we will see, in the $\frac{SO(4)}{SO(2)\times SO(2)}$ case this does not lead to additional restrictions. 

In the present case, the scalar function has the form
\bea
{\tilde B}(y)=e^{i{n}_2(\alpha+\beta)}{Y}_2(y_2),
\eea
and $Y_2(y_2)$ satisfies the second equation (\ref{SO422ODE}):
\bea
&&\frac{1}{(1-y_2)^{\sigma}}
\frac{d}{dy_2}\left[(1-y_2)^{\sigma}(1+y_2) \frac{dY_2}{dy_2}\right]-
\frac{n_2^2}{1+y_2}Y_2+\frac{{\tilde\lambda}_{scalar}}{1-y_2} Y_2=0.\nonumber
\eea
The vector field $C_i$ has the form 
\bea\label{VecCsum}
C_i dx^i=e^{i n_1(\alpha-\beta)}\left[V_1 dy_1+
{V}_-(d\alpha-d\beta)\right].
\eea
where $V_1$ and ${V}_-$ are functions of $y_1$. Substitution into the second equation in (\ref{VectBranch1ODE}) gives and expression for $V_1$,
\bea\label{V1SO422Brnch1}
V_1=\frac{in_1(1-y_1)V_-'}{\lambda_{vec}(1+y_1)-n_1^2(1-y_1)}\,,
\eea
as well as a differential equation for $V_-$:
\bea\label{SO44VecEqn}
\frac{1+y_1}{(1-y_1)^\sigma}\frac{d}{dy_1}\left[
\frac{(1-y_1)^{\sigma+1}(1+y_1)V_-'}{\lambda_{vec}(1+y_1)-n_1^2(1-y_1)}\right]+V_-=0
\eea
The Lorenz condition (\ref{LorentzOne}), 
\bea
\d_i\left[e^{-2\sigma\phi}\sqrt{g}g^{ij}{C}_{j}\right]=0,\nonumber
\eea
is automatically satisfied, and the eigenvalues of the problem (\ref{VecHelm1}) are given by 
\bea
\Lambda={{\tilde \lambda}_{scalar}+\lambda_{vector}}\,.
\eea
Interstingly, (\ref{SO44VecEqn}) and the first equation in (\ref{SO422ODE}) have the same set of eigenvalues, and solutions $V_-$ can be written in terms of eigenfunctions $Y_1$ by
\bea\label{VminusBr1SO422}
V_-= (1+y_1)\frac{dY_1}{dy_1}-\frac{(n_1)^2\sigma}{\lambda}Y_1,\quad \lambda_{vec}=\lambda_{scalar}
\eea
\item {\bf Vector fields on the $y$--space:}\\
This situation is analogous to the case (a) with a replacement 
\bea
x\rightarrow y,\quad {\tilde \lambda}_{scalar}\rightarrow {\lambda}_{scalar},\quad
{\lambda}_{vec}\rightarrow {\tilde\lambda}_{vec}
\eea  
The ansatz for the vector field is
\bea
A={B}(x) {\tilde C}_a(y) dy^a\,.
\eea
with 
\bea
{B}(x)=e^{i{n}_1(\alpha-\beta)}{Y}_1(y_1),\quad
{\tilde C}_a dx^a=e^{i n_2(\alpha+\beta)}\left[V_2 dy_2+
{V}_+(d\alpha+d\beta)\right].
\eea
Function $Y_1(y_1)$ satisfies the first ODE from (\ref{SO422ODE}),
\bea
\frac{1}{(1-y_1)^{\sigma}}
\frac{d}{dy_1}\left[(1-y_1)^{\sigma}(1+y_1) \frac{dY_1}{dy_1}\right]-
\frac{n_1^2}{1+y_1}Y_1+\frac{\lambda_{scalar}}{1-y_1} Y_1=0\,,
\eea
and $V_+$ satisfies a counterpart of (\ref{SO44VecEqn})
\bea\label{VecSO422e2}
\frac{1+y_2}{(1-y_2)^\sigma}\frac{d}{dy_2}\left[
\frac{(1-y_2)^{\sigma+1}(1+y_2)V_+'}{{\tilde\lambda}_{vec}(1+y_2)-n_2^2(1-y_2)}\right]+V_+=0
\eea
As in the case (a), the sets of scalar and vector eigenvalues, $\{{\tilde\lambda}_{scalar}\}$ and 
$\{{\tilde\lambda}_{vec}\}$, are the same, and the eigenfunctions of (\ref{VecSO422e2}) and (\ref{SO422ODE}) are related by
\bea
V_+= (1+y_2)\frac{dY_2}{dy_2}-\frac{(n_2)^2\sigma}{{\tilde\lambda}_{vect}}Y_2,\quad 
\lambda_{vec}=\lambda_{scalar}
\eea
Function $V_2$ is given by 
\bea
V_2=\frac{in_2(1-y_2)V_+'}{{\tilde\lambda}_{vec}(1+y_2)-n_2^2(1-y_2)}\,,
\eea
and the eigenvalues of the problem (\ref{VecHelm1}) are 
\bea
\Lambda=\frac{2\pi}{k}({{\lambda}_{scalar}+\tilde\lambda_{scalar}})\,.
\eea
\item {\bf The scalar mode:}\\
The ansatz for the gauge field is given by (\ref{VectScalBranch})--(\ref{VectScalBranchEqn2}):
\bea
A={\tilde B}(y) d B(x)+\mu {B}(x) d{\tilde B}(y), 
\eea
and in the present case, 
\bea
{B}(x)=e^{i{n}_1(\alpha-\beta)}{Y}_1(y_1),\quad 
{\tilde B}(y)=e^{i{n}_2(\alpha+\beta)}{Y}_2(y_2).
\eea
Functions $Y_1$ and $Y_2$ satisfy equations (\ref{SO422ODE}), and the eigenvalue $\Lambda$ and parameter $\mu$ are given by (\ref{LamMuVec})
\bea
\Lambda=\frac{2\pi}{k}({{\lambda}_{1}+{\lambda}_{2}})\,,\quad
\mu=-\frac{{\lambda}_{1}}{{\lambda}_{2}}\,.
\eea
\end{enumerate}
To summarize, application of the separable ansatz (\ref{VecSeparAnstz}) to 
$\frac{SO(4)}{SO(2)\times SO(2)}$ describes three physical degrees of freedom per each pair of eigenvalues $(\lambda_1,\lambda_2)$ of the system (\ref{SO422ODE}). The full spectrum describes three copies of (\ref{LambSO422}) corresponding to cases (a), (b) and (c). 

Our analysis was based on the product structure of the space (\ref{SO422ProdStruct}), but it is also instructive to compare with the ansatz (\ref{VectAnstz}) inspired by Maxwell's equation on black hole geometries. To do so, we write the metric (\ref{SU2tmSU2}) in terms of frames:
\bea
&&\hskip -0.5cm ds^2=\frac{k}{2\pi}(e^{1+}_\mu e^{1-}_\nu dx^\mu dx^\nu+e^{2+}_\mu e^{2-}_\nu dx^\mu dx^\nu),\nn
&&\hskip -0.5cm e^{1\pm}_\mu dx^\mu=\sqrt{\frac{{1}}{1-y_1^2}}\left[dy_1\pm i(1+y_1)(d\alpha+d\beta)\right],\nn
&&\hskip -0.5cm e^{2\pm}_\mu dx^\mu=\sqrt{\frac{{1}}{1-y_2^2}}\left[dy_2\pm i(1+y_2)(d\alpha-d\beta)\right],\\
&&\hskip -0.5cm e_{1\pm}^\mu \d_\mu=\sqrt{{1-y_1^2}}\left[\d_{y_1}\mp \frac{i}{(1+y_1)}\d_{\alpha+\beta}\right],\quad
e_{2\pm}^\mu \d_\mu=\sqrt{{1-y_2^2}{}}\left[\d_{y_2}\mp \frac{i}{(1+y_2)}\d_{\alpha-\beta}\right]\nonumber
\eea
Then equations (\ref{VectSeparOne1}), (\ref{VecCsum}), (\ref{V1SO422Brnch1}), (\ref{VminusBr1SO422}), lead to simple expressions for the projections:
\bea\label{Vec422optA}
\mbox{(a)}:&& e_{1+}^\mu A_\mu=-\frac{i(\lambda_1-n_1\sigma)}{\lambda_1}e_{1+}^\mu \d_\mu Z,\quad
e_{1-}^\mu A_\mu=\frac{i(\lambda_1+n_1\sigma)}{\lambda_1}e_{1-}^\mu \d_\mu Z,\nn
&&e_{2\pm}^\mu A_\mu=0,\quad Z=e^{in_1(\alpha-\beta)+in_2(\alpha+\beta)}Y_1(y_1)Y_2(y_2).
\eea
Here we used the first equation form (\ref{SO422ODE}) to eliminate higher derivatives of $Y_1(y_1)$. Similarly, for the other branches we find:
\bea\label{Vec422optB}
\mbox{(b)}:&& e_{2+}^\mu A_\mu=-\frac{i(\lambda_2-n_2\sigma)}{\lambda_2}e_{2+}^\mu \d_\mu Z,\quad
e_{2-}^\mu A_\mu=\frac{i(\lambda_2+n_2\sigma)}{\lambda_2}e_{2-}^\mu \d_\mu Z,\nn
&&e_{1\pm}^\mu A_\mu=0,\quad Z=e^{in_1(\alpha-\beta)+in_2(\alpha+\beta)}Y_1(y_1)Y_2(y_2);\\
\mbox{(c)}:&& e_{1\pm}^\mu A_\mu=e_{1\pm}^\mu \d_\mu Z,\quad
e_{2\pm}^\mu A_\mu=-\frac{\la_1}{\la_2}e_{2\pm}^\mu \d_\mu Z,\quad 
Z=e^{in_1(\alpha-\beta)+ in_2(\alpha+\beta)}Y_1(y_1)Y_2(y_2).\nonumber
\eea
All three cases, as well as their arbitrary linear combinations, match the structure (\ref{VectAnstz})
\bea\label{VectAnstzO422}
&&e_{1\pm}^\mu A_\mu=b_{1\pm} e_{1\pm}^\mu\d_\mu {\tilde Z},\quad 
e_{2\pm}^\mu A_\mu=b_{2\pm} e_{2\pm}^\mu\d_\mu {\tilde Z}
\eea
with constant coefficients $(b_{1\pm},b_{2\pm})$.

\bigskip

To summarize, in this subsection we demonstrated a full separation of variables in the scalar and vector equations on the background of the WZW model for the $SO(4)/[SO(2)\times SO(2)]$ coset. We found that, up to an extra degeneracy in the vector sector, the scalar and vector spectra are identical and the eigenvalues are given by
\bea
\Lambda=\frac{2\pi}{k}(\lambda_1+\lambda_2),\quad 
\lambda_{1,2}=j_{1,2}(j_{1,2}+1)-n_{1,2}^2.
\eea
The components of the vector field are expressed in terms of the scalar by one of the options (\ref{Vec422optA})--(\ref{Vec422optB}), and various ingredients of the scalar eigenfunction (\ref{ScalO422tmp}) satisfy ordinary differential equations (\ref{SO422ODE}).

\subsection{Gauging and T--duality}
\label{SecTdual}

In this section we have analyzed the eigenvalues problems for scalar and vector fields on the backgrounds of the (gauged) WZW models corresponding to $SO(4)$ and its cosets, $SO(4)/H$. Although the differential equations describing the dynamical excitations varied with the subgroup $H$, there were some similarities between them, and in this subsection we will address the origin of these similarities. Specifically, we will demonstrate that the target spaces of various $SO(4)/H$ are related to each other by T duality, and that equations for excitations transform under such dualities in a simple way.

We begin with the $SO(4)$ WZW model that produces the geometry (\ref{MetrSO4y}). Defining new coordinates $(\alpha_\pm,\beta_\pm)$ by
\bea 
\alpha_\pm=\frac{(\alpha_L-\beta_L)\pm (\alpha_R-\beta_R)}{2},\quad \beta_\pm=\frac{(\alpha_++\beta_L)\pm(\alpha_L-\beta_L)}{2},
\eea 
we can write the $B$ field and the the angular parts of the metric as 
\bea \label{tempTdual}
&&ds^2=(1+y_1)d\alpha_+^2+(1-y_1)d\alpha_-^2+(1+y_2)d\beta_+^2+(1-y_2)d\beta_-^2,\nn
&&B=(1-y_2)d\beta_+\wedge d\beta_-+(1-y_1)d\alpha_+\wedge d\alpha_-.
\eea 
To simplify the discussion, we rescaled the metric and the $B$--field by the factor $\frac{k}{2\pi}$. Performing T--dualities the $\alpha_-$ and $\beta_-$ directions, one finds a new background with the metric 
\bea 
ds^2=\frac{1+y_1}{2(1-y_1)}d\alpha_-^2+\frac{1}{2}(d\alpha_--2d\alpha_+)^2+\frac{1+y_2}{2(1-y_2)}d\beta_-^2+\frac{1}{2}(d\beta_--2d\beta_+)^2,
\eea 
but without the $B$ field. Comparison with (\ref{SU2tmSU2}) shows that the dual dual geometry is $[SO(4)/[SO(2)\times SO(2)]]\times U(1)^2$. This agrees with a general statement that gauging of any  $SO(2)$ symmetry is equivalent to a T duality \cite{Tduality}. By performing only one T--duality in (\ref{tempTdual}), one would find $[SO(4)/SO(2)]\times U(1)$. 

The map between $SO(4)/SO(2)$ and $SO(4)/[SO(2)\times SO(2)]$ cosets is slightly more interesting. The T--duality corresponding to this map is performed along some combination of angles appearing in (\ref{SO4SO2}). Specifically, introducing a new coordinate $\tau=2\gamma+\alpha$, we can rewrite the $SO(4)/SO(2)$ metric (\ref{SO4SO2}) as\footnote{In the subsection we have dropped the factor $k/\pi$.}:
\bea 
ds^2&=&\frac{(y_1y_2-1)}{y_1+y_2-2}d\alpha^2+\frac{2(y_1-y_2)}{y_1+y_2-2}d\alpha d\beta-\frac{(y_1-1)(y_2-1)}{y_1+y_2-2}d\tau^2\nn
&&-\frac{y_1+y_2+2}{y_1+y_2-2}d\beta^2+\frac{dy_1^2}{2-2y_1^2}+\frac{dy_2^2}{2-2y_2^2}\nn
B&=&d\tau\wedge \left[d\beta \frac{y_2-y_1}{y_2+y_1-2}-d\alpha\frac{(y_1y_2-1)}{y_2+y_1-2}\right],\qquad
e^{-2\Phi}=2-y_1-y_2
\eea 
T duality along $\tau$ direction removes the $B$--field and makes the dilaton separable:
\bea\label{SO4SO2dual}
ds^2&=&\frac{1-y_1y_2}{(y_1-1)(y_2-1)}[d\alpha^2+d\beta^2]-\frac{2(y_1-y_2)d\alpha d\beta}{(y_1-1)(y_2-1)}-\frac{y_1+y_2-2}{(y_1-1)(y_2-1)}d\tau^2\nn
&&2d\tau\frac{(y_1-y_2)d\beta-(1-y_1y_2)d\alpha}{(y_1-1)(y_2-1)}+\frac{dy_1^2}{2-2y_1^2}+\frac{dy_2^2}{2-2y_2^2}\nn
e^{-2\Phi}&=&(y_1-1)(y_2-1)
\eea
An additional shift, $\alpha\rightarrow\alpha+\tau$ leads to a simpler metric (\ref{SU2tmSU2}) with an additional flat direction $\tau$:
\bea
ds^2&=&\frac{1-y_1y_2}{(y_1-1)(y_2-1)}[d\alpha^2+d\beta^2]-\frac{2(y_1-y_2)d\alpha d\beta}{(y_1-1)(y_2-1)}+
d\tau^2+\frac{dy_1^2}{2-2y_1^2}+
\frac{dy_2^2}{2-2y_2^2}\nn
&=&\frac{1+y_2}{2(1-y_2)}[d\alpha+d\beta]^2+\frac{1+y_1}{2(1-y_1)}[d\alpha-d\beta]^2+
d\tau^2+\frac{dy_1^2}{2-2y_1^2}+
\frac{dy_2^2}{2-2y_2^2}\nn
e^{-2\Phi}&=&(y_1-1)(y_2-1)
\eea
As expected, this is the $[SU(2)/U(1)]\times [SU(2)/U(1)]\times U(1)$ geometry.

Once various $SO(4)/H$ backgrounds are shown to be related by T--dualities, separation of variables on one of them guarantees separation on another provided that dynamical equations remain invariant. In particular, the scalar equation (\ref{Helpm}) is invariant under a T--duality if and only if $\sigma=1$, so separability of the Helmholtz equation on $SO(4)$, where the dilaton is trivial, would imply separability on $SO(4)/H$ only for $\sigma=1$. We saw this explicitly for the $SO(4)/SO(2)$ coset in section \ref{SecSO42}. Interestingly, the scalar equation on the $SO(4)/[SO(2)\times SO(2)]$ geometry separates for an arbitrary $\sigma$ (see (\ref{SO422ODE})), but such ``bonus separation'' is not a consequence of T--duality. 

To separate the vector equation (\ref{VecEqnSO42}), one needs to build special frames associated with the Killing--Yano tensors (see, for example, (\ref{KYTso42})). The behavior of 
the Killing--Yano tensors (KYT) under T-duality was studied in \cite{ChLKil}, where it was shown that while the ordinary KYTs may disappear, the twisted KYTs are preserved. Interestingly, it is precisely such twisted Killing--Yano tensors,  (\ref{TwistKY1}) and (\ref{TwistKY2}), that are responsible for separation of the vector equations after T--duality. Therefore, we have demonstrated that separations of the scalar and vector equations on the $SO(4)/H$ cosets are not accidental, but rather they are guaranteed by the relation between gauging and T--duality \cite{Tduality} and by the transformation of dynamical equations and Killing--Yano tensors under the duality \cite{ChLKil}.

\section{The $SO(5)$ sigma model}
\label{SecSO5}

In this section we will look at separation of variables in the $SO(5)$ sigma model. Unfortunately the full separation of variables encountered in previous section for $SO(4)$ and it cosets does not persist for $SO(5)$, but we find several interesting sectors which admit a partial separation. We begin with reviewing parameterization of $SO(5)$ and an algebraic construction of the scalar eigenfunctions developed in \cite{LTian}. In section \ref{SecSO5group} we also present some simple examples of wavefunctions which inspire the analysis in the rest of the discussion of the $SO(5)$ group. In sections \ref{SecSubNoX} and \ref{SecSubXlinear} we construct two infinite classes of separable eigenfunctions by solving the Helmholtz equation. Each family is parameterized by four discrete quantum numbers.  In section \ref{SecSubSymRep} we use an algebraic procedure to construct additional infinite families of separable solutions which depend on four parameters as well. Finally, in section \ref{SecSubSpehere} we discuss partial separation for a different parameterization of $SO(5)$ as well as its extensions to larger groups.

\bigskip

The action of the $SO(5)$ WZW model,
\bea\label{WZWSO5}
S=-\frac{k}{2\pi}\int d^2\sigma \eta^{\alpha\beta}\mbox{tr}(g^{-1}\d_\alpha g g^{-1}\d_\beta g)+
\frac{ik}{6\pi}\int \mbox{tr}(g^{-1}d g\wedge  g^{-1}d g\wedge  g^{-1}d g)
\eea
is invariant under the $SO(5)_L\times SO(5)_R$ global symmetry. Since $SO(5)$ has rank two, the sigma model (\ref{WZWSO5}) has $2+2=4$ commuting Killing vectors. It is useful to realize these $U(1)$ symmetries by simple translations, and this can be accomplished by the following parameterization of the group element $g$:
\bea\label{SO5Param}
\hskip -0.5cm
g=h[\alpha_L,\beta_L]\left[\begin{array}{cc}
I-B_X X^T X&-B_XX^T\\
B_X X&I-B_X X X^T
\end{array}\right]
\left[\begin{array}{ccc}
I-B_Y YY^T&B_Y Y&0\\
-Y^T B_Y&I-Y^T B_Y Y&0\\
0&0&1
\end{array}\right]h[\alpha_R,\beta_R]\,.\hskip -1.5cm\nn
\eea
Here vector $X$, scalar $B_X$, and matrices $(Y,B_Y)$, are defined by 
\bea
X=(X_1,X_2,X_3,X_4),\quad 
B_X=\frac{2}{1+XX^T}\,\quad
Y=\mbox{diag}(Y_1,Y_2),\quad 
B_Y=\frac{2}{1+YY^T}\,.
\eea
We also defined $h[\alpha,\beta]$ as a matrix function of two angles:
\bea
h[\alpha,\beta]=\left[\begin{array}{ccccc}
c_{\alpha}&s_{\alpha}&0&0&0\\
-s_{\alpha}&c_{\alpha}&0&0&0\\
0&0&c_{\beta}&s_{\beta}&0\\
0&0&-s_{\beta}&c_{\beta}&0\\
0&0&0&0&1
\end{array}\right]\,.
\eea
Note that matrix $h_L=h[\alpha_L,\beta_L]$ appears in the action (\ref{WZWSO5})
only in the combination $h^{-1}_L dh_L$, so coordinates $(\alpha_L,\beta_L)$ are cyclic. Similarly, 
matrix $h_R=h[\alpha_R,\beta_R]$ appears only in the combination $dh_R h_R^{-1}$, so coordinates $(\alpha_R,\beta_R)$ are cyclic as well. The full metric corresponding to (\ref{WZWSO5}) is rather complicated, and here we just stress one important property, which is easy to verify. If we write $X_i=R \mu_i$, where three variables $\mu_i$ are subject to constraint $\sum\mu_1^2=1$, then 
\bea\label{Rsplit}
ds^2=\frac{k}{2\pi}\left[\frac{8 dR^2}{(1+R^2)^2}+(\mbox{terms without }dR)\right]
\eea
In other words, the cross terms between $dX_i$ and remaining coordinates can be written in terms of $d\mu_i$.

In this section we will study the scalar equation 
\bea\label{HelpmSO5}
\nabla^2\Phi=-\frac{\pi}{k}\Lambda\Phi
\eea
in the geometry (\ref{WZWSO5})--(\ref{SO5Param}). As demonstrated in \cite{PS}, the eigenvalues of this equation can be expressed in terms of the quadratic Casimir of the gauge group, and the $SO(5)$ case, 
the result is 
\bea\label{SO5eigVal}
\Lambda=l_1(l_1+3)+l_2(l_2+1),\quad l_1\ge l_2,
\eea
where $(l_1,l_2)$ are either both integers or both half--integers. Our goal is to construct the corresponding eigenfunctions. In contrast to the situation described in the previous section, equation (\ref{HelpmSO5}) is not fully separable for the $SO(5)$ WZW model, but there are several separable families and they will be described in separate subsections. The simplest family follows from the observation (\ref{Rsplit}): if we assume that $\Phi$ is a function of $R$ only, then the equation (\ref{HelpmSO5}) becomes\footnote{We used the expression for the determinant of the metric.} 
\bea\label{tempAlabel}
\frac{(1+R^2)^4}{4R^3}\frac{d}{dR}\left[\frac{R^3}{(1+R^2)^2}\frac{d\Phi}{dR}\right]+\Lambda\Phi=0
\eea
The normalizable solutions are 
\bea\label{NeutralHyper}
\Phi=F\left[-k,3+k;2;\frac{1}{1+R^2}\right],\quad \Lambda=\frac{(3+2k)^2-9}{4}\,,
\eea
where $k$ is a non--negative integer, so we recover (\ref{SO5eigVal}) with $(l_1,l_2)=(k,0)$. In the remaining part of this section we will extend the explicit solution (\ref{NeutralHyper}) to more general families.

\subsection{Eigenfunctions from group theory}
\label{SecSO5group}

Before analyzing differential equations, it is useful to recall the algebraic construction for the eigenfunctions of the Helmholtz equation (\ref{HelpmSO5}). As demonstrated in \cite{PS}, all such eigenfunctions can be constructed as polynomials in the matrix elements of $g$. Specifically, each eigenvalue (\ref{SO5eigVal}) corresponds to an irreducible representation of $SO(5)$. Such representations are characterized by Young tableaux, which in turn specify representations of the permutation group $S_5$. Then the wavefunction 
$\Phi$ is written as the sum over relevant permulations $P$ \cite{LTian}
\bea\label{PhiYoung}
\Phi=\sum_P (-1)^{\sigma(P)} g_{i_1j_{P[1]}}\dots g_{i_Lj_{P[L]}}-(traces)
\eea
The wavefunction is fully specified by the set of $2L$ indices $(i_1,\dots,i_L,j_1,\dots j_L)$ and the signatures $\sigma(P)$ associated with the Young tableau. In this subsection we will present several examples of eigenfunctions (\ref{PhiYoung}) for representation of $SO(5)$ with small $L$, and in the subsequent subsections the patterns observed in these examples will be used to construct infinite separable families. 

\bigskip

The first set of states corresponds to the Young tableau with one box. The eigenvalue is 
\bea
\Lambda=4,
\eea 
and the eigenfunctions are arbitrary linear combinations of the matrix elements $g_{ij}$. To make the $[U(1)]^4$ symmetries explicit, we focus on the combinations which have specific charges under these transformations. There are $25$ states in total. One of them is neutral, and it corresponds to (\ref{NeutralHyper}) with $k=1$:
\bea\label{SO5rFunc}
\frac{1-R^2}{1+R^2}
\eea
This is the only state in the $k=1$ representation that does not have angular or $X_a$ dependence. 

To write the remaining states in the $k=1$ representation, it is convenient to introduce three combinations of the coordinates $(R,Y_1,Y_2)$,
\bea\label{SO5denom}
D=\frac{(1+Y_1^2)(1+Y_2^2)(1+R^2)}{1-Y_1^2Y_2^2},\quad
y_+=\frac{Y_1+Y_2}{1-Y_1 Y_2},\quad y_-=\frac{Y_1-Y_2}{1+Y_1 Y_2},
\eea
as well as six complex combinations of $X_a$,
\bea\label{SO5ComplCord}
&&z_1=X_1+i X_2,\quad z_2=X_3+i X_4,\\
&&Z_{1+}={z}_1-Y_+{z}_2,\quad Z_{2+}={z}_2+y_+{z}_1,\quad Z_{1-}={z}_1-y_-{\bar z}_2,\quad
Z_{2-}={z}_2+y_-{\bar z}_1\,.\nonumber
\eea
Then we find that the $25$ states in the $k=1$ representation can be divided in four groups:
\begin{enumerate}[1.]
\item One state (\ref{SO5rFunc}) without angular or $X_a$ dependence.
\item
Eight states are linear in $X_a$. They are given by
\bea\label{eqn126}
\frac{e^{-2i\alpha_L}{z}_1}{1+R^2},\quad
 \frac{e^{-2i\beta_L}{z}_2}{1+R^2},\quad
\frac{e^{2i\alpha_R}(Z_{1+}-Y_-{\bar Z}_{2+})}{D},\ 
\frac{e^{2i\beta_R}(Z_{2+}+Y_-{\bar Z}_{1+})}{D},
\eea
and their complex conjugates. 
\item
Eight states charged under $U(1)_{\alpha_L}$ symmetry are given by
\bea\label{eqn127}
&&
e^{-2i(\alpha_L+\alpha_R)}\frac{(1+Z_{2+}{\bar Z}_{2-})}{D},\
e^{-2i(\alpha_L+\beta_R)}\frac{(Z_{1+}{\bar Z}_{2-}-Y_+)}{D},\nn 
&&e^{2i(\alpha_R-\alpha_L)}\frac{(Y_+Y_-+{ Z}_{1+}{Z}_{1-})}{D},\
e^{-2i(\alpha_L-\beta_R)}\frac{(Z_{1-}{Z}_{2+}-Y_-)}{D},
\eea
and their complex conjugates.
\item
Eight states charged under $U(1)_{\beta_L}$ symmetry can be obtained from (\ref{eqn127}) and their complex conjugates by the replacements
\bea
\alpha_L\rightarrow \beta_L,\quad \alpha_R\rightarrow\beta_R,\quad 
Z_{1\pm}\rightarrow Z_{2\pm},\quad Z_{2\pm}\rightarrow -Z_{1\pm}.
\eea
\end{enumerate}
The extensions of these groups to general families will be discussed in section \ref{SecSubXlinear}. We conclude this subsection by listing some solutions corresponding to antisymmetric representation characterized by a Young tableau with two boxes. The eigenvalue 
$\Lambda=4$ has degenecy $100$, and the wavefunctions are specified by two antisymmetric pairs of indices, $(i,j)$ and $(k,l)$.\footnote{Recall that $g$ is a $5\times 5$ matrix, so the infices $(i,j,k,l)$ range from one to five. Then the antisymmetric combinations, $(i,j)$ and $(k,l)$, can take $10$ possible values each. This explains the $100$--fold degeneracy of the eigenvalue $\Lambda=6$.} Up to a nomalization factor, the wavefunctions are given by
\bea\label{SO5Anti2box}
\Phi_{ij;kl}=g_{ik}g_{jl}-g_{il}g_{jk}\,. 
\eea
As we saw already in the case of the fundmantal representation, it is conveninent to introduce complex coordinates (\ref{SO5ComplCord}), so we will use the values $(z_1,{\bar z}_1,z_2,{\bar z}_2,5)$ for indices $(i,j,k,l)$ as well. For example,
\bea
\Phi_{z_1j;kl}=\Phi_{1j;kl}+i\Phi_{2j;kl},\quad \Phi_{z_2j;kl}=\Phi_{1j;kl}-i\Phi_{2j;kl}
\eea
Substituting the explicit expressions for the matrix elements of $g$, we observe that the following combinations, as well as their complex conjugates, depend on $(z_1,{\bar z}_1,z_2,{\bar z}_2)$ only through $R^2$:
\bea\label{YsolnEx}
&&\Phi_{z_1z_2;z_1z_2}=\frac{4E_{1,1,1,1}(1+Y_1Y_2)}{D(1-Y_1Y_2)}\,,\quad
\Phi_{z_1z_2;{\bar z}_1{\bar z}_2}=
\frac{4E_{1,1,-1,-1}(Y_1-Y_2)^2}{D[1-(Y_1Y_2)^2]}\nn
&&\Phi_{z_1{\bar z}_2;z_1{\bar z}_2}=\frac{4E_{1,-1,1,-1}(1-Y_1Y_2)}{D(1+Y_1Y_2)}\,,\quad
\Phi_{z_1{\bar z}_2;{\bar z}_1{z}_2}=
\frac{4E_{1,-1,-1,1}(Y_1+Y_2)^2}{D[1-(Y_1Y_2)^2]}
\eea
Here we introduced a convenient shorthand notation
\bea
E_{a,b,c,d}=e^{-2i(a\alpha_L+b\beta_L+c\alpha_R+d\beta_R)}
\eea
Note that wavefunctions (\ref{YsolnEx}) have a separable structure
\bea\label{YsolnExStruc}
\Phi=e^{-2i(a\alpha_L+b\beta_L+c\alpha_R+d\beta_R)}f(R)g(Y_1,Y_2)\,.
\eea
In the next two subsections we will construct the most general function of the form (\ref{YsolnExStruc}) that solves the scalar equation (\ref{HelpmSO5}). In sections \ref{SecSubXlinear} and \ref{SecSubSymRep} extensions to several classes of 
$z$--dependent solutions will be discussed as well, and they will contain the states (\ref{eqn126}) and (\ref{eqn127}) as special cases.

\subsection{Factorization of the $R$ dependence}
\label{SecDressR}

Before finding the most general solutions of the form (\ref{YsolnExStruc}), it is instructive to start with a specific solution, such as one of the functions listed in (\ref{YsolnEx}), and explore the possibility of changing function $f(R)$ while keeping $g(Y_1,Y_2)$ and constants $(a,b,c,d)$ fixed. This subsection is dedicated to the discussion of such ``$R$--dressing'', and our starting point will be slightly more general than (\ref{YsolnExStruc}).

\bigskip

Let us assume that the Helmholtz equation (\ref{HelpmSO5}) has a solution of the form 
\bea\label{SolnToDress}
\Phi_{\Lambda_0}=\frac{R^p}{(1+R^2)^q}\,\tilde\Phi(\gamma_i,Y_a,\frac{X_j}{R}),\quad 
\gamma_i=\{\alpha_L,\beta_L,\alpha_R,\beta_R\}
\eea
In particular, wavefunctions (\ref{YsolnExStruc}) and (\ref{eqn126}) fit this pattern. We will now demonstrate that equation (\ref{HelpmSO5}) admits a family of normalizable solutions which are obtained by ``dressing'' solutions (\ref{SolnToDress}) by some specific function of the radial coordinate: 
\bea\label{SolnToDressGss}
\Phi^{(k)}_{\Lambda_k}=\frac{R^p}{(1+R^2)^q}\,f_{k,p,q}(R)\tilde\Phi(\gamma_i,Y_a,\frac{X_j}{R}),\quad
\Lambda_k=\Lambda_0+(2k+2q+3)^2-(2q+3)^2\,.
\eea
The ``dressed'' solution depends on an integer parameter $k$. To prove (\ref{SolnToDressGss}), we recall that the metric has the form (\ref{Rsplit}), where ``terms without $dR$'' contain $(d\gamma_i,dY_a,d\mu_j)$, where $\mu_j$ are three angles from a constrained set of four parameters:
\bea
X_j=R \mu_j, \qquad \sum \mu_j^2=1.
\eea
The differential equation (\ref{HelpmSO5}) has the form
\bea\label{DiffEqnRfact}
\frac{(1+R^2)^4}{4R^3}\frac{\d}{\d R}\left[\frac{R^3}{(1+R^2)^2}\frac{\d\Phi}{\d R}\right]+{\tilde\nabla}^2\Phi+\Lambda\Phi=0,
\eea
where ${\tilde\nabla}^2$ has a complicated $R$--dependence, but no $R$--derivatives. Writing equation (\ref{DiffEqnRfact}) for two wavefunctions, (\ref{SolnToDressGss}) and (\ref{SolnToDress}), and combining the results to eliminate the terms with ${\tilde\nabla}^2$, we find
\bea\label{tempDec23}
&&\frac{(1+R^2)^4}{4R^3}\frac{\d}{\d R}\left[\frac{R^3}{(1+R^2)^2}\frac{\d\Phi^{(k)}_{\Lambda_k}}{\d R}\right]+(\Lambda_k-\Lambda_0)\Phi^{(k)}_{\Lambda_k}\\
&&\qquad=\frac{(1+R^2)^{4+q}}{4R^{3+p}}\frac{\d}{\d R}\left[\frac{R^3}{(1+R^2)^2}\frac{d}{d R}\frac{R^p}{(1+R^2)^q}\right]\Phi^{(k)}_{\Lambda_k}\,.\nonumber
\eea
To simplify this equation, we define a new function $h$ by
\bea
h\left[\frac{R^2}{1+R^2}\right]\equiv f_{k,p,q}(R).
\eea
Then equation (\ref{tempDec23}) becomes
\bea
x(1-x)h''+[2+p-2(2+q)x]h'+(\Lambda_k-\Lambda_0)h=0,
\eea
and the solution regular at $R=0$ can be expressed in terms of the hypergeometric function:
\bea
&&f_{k,p,q}(R)=F\left[-k,3+k+2q;2+p;\frac{R^2}{1+R^2}\right],\\
&&\qquad\mbox{where}\quad
\Lambda_k=\Lambda_0+\frac{(2k+2q+3)^2-(2q+3)^2}{4}\,.\nonumber
\eea
Normalizability at large values of $R$ requires $k$ to be a non--negative integer. 

To summarze, we have demonstrated that every wavefunction of the form (\ref{SolnToDress}) gives rise to a one--parametric family of normalizable solutions of the Helmholtz equation (\ref{HelpmSO5}). The wavefunctions are
\bea\label{SolnDressed}
\Phi^{(k)}_{\Lambda_k}=\frac{R^p}{(1+R^2)^q}\, F\left[-k,3+k+2q;2+p;\frac{R^2}{1+R^2}\right]
\tilde\Phi(\gamma_i,Y_a,\frac{X_j}{R}),
\eea
and the eigenvalues are
\bea\label{SolnDressedLam}
\Lambda_k=\Lambda_0+\frac{(2k+2q+3)^2-(2q+3)^2}{4}\,.
\eea
Normalizabilty requires $k$ to be a non--negative integer. An alternative form of (\ref{SolnDressed})\footnote{We dropped a constant multiplicative factor.},
\bea\label{SolnDressedAlt}
\Phi^{(k)}_{\Lambda_k}=\frac{R^p}{(1+R^2)^q}\, F\left[-k,3+k+2q;2+2q-p;\frac{1}{1+R^2}\right]
\tilde\Phi(\gamma_i,Y_a,\frac{X_j}{R}),
\eea
may be useful as well. In particular, the solution (\ref{NeutralHyper}) is recovered by choosing the trivial function $\tilde\Phi$ and $p=q=0$.

We conclude this subsection with presenting an example of the dressing (\ref{SolnDressed}). Observing that 
the wavefunctions (\ref{YsolnEx}) have $(p,q)=(0,1)$, we can dress the first wavefunction as 
\bea
\Phi^{(k)}_{z_1z_2;z_1z_2}=\frac{4E_{1,1,1,1}(1+Y_1Y_2)}{D(1-Y_1Y_2)}
F\left[-k,5+k;2;\frac{R^2}{1+R^2}\right]\,,\ \Lambda_k=\Lambda_0+\frac{(2k+5)^2-25}{4}\,.\nn
\eea
The remaining wavefunction from (\ref{YsolnEx}), as well as examples from (\ref{eqn126}) can be dressed in the same way.

\subsection{Separable $X$--independent solutions}
\label{SecSubNoX}

In this subsection we will generalize the solutions (\ref{YsolnEx}) to wavefunctions which have the form (\ref{YsolnExStruc}). For fixed function $g$ and parameters $(a,b,c,d)$, solution (\ref{YsolnExStruc}) covers a one--parameter family of ``dressed'' wavefunctions analyzed in the previous subsection. To avoid unnecessary complications associated with $f(R)$, here we will focus on the ``seed solutions'' (\ref{SolnToDress}):
\bea\label{YsolnStart}
\Phi=e^{2i[n_1\alpha_L+n_2\beta_L+n_3\alpha_R+n_4\beta_R]}\frac{g[Y_1,Y_2]}{(1+R^2)^q}\,,
\eea
and the ``dressing'' will be added in the end. 
In contrast to (\ref{SolnToDress}), equation (\ref{YsolnStart}) lists the $[U(1)]^4$ charges explicitly. Also, since we are looking for solutions independent of $\mu_j=X_j/R$, the parameter $p$ in the seed solution (\ref{SolnToDress}) vanishes. Substitution of the ansatz (\ref{YsolnStart}) into the Helmholtz equation (\ref{HelpmSO5}) leads to a complicated overdetermined system of equations for the function $g[Y_1,Y_2]$.\footnote{Specifically, variables $(X_1,X_2,X_3,X_4)$  appear in the equation (\ref{HelpmSO5}) in various combinations, not only as $R^2$. This leads to a system of PDEs for one function $g[Y_1,Y_2]$.} The explicit form of these equations is not very illuminating, so we will present only the logic for solving them.
\begin{enumerate}[(i)]
\item Once the ansatz (\ref{YsolnStart}) is substituted into the equation (\ref{HelpmSO5}), one finds an equation that contains various functions of $(X_1,X_2,X_3,X_4)$ and $R$. Expressing $X_1$ in terms of the remaining variables, one finds a system with {\it independent} $(R,X_2,X_3,X_4)$. In particular, the coefficient in front of the product $(X_2 X_4)$ contains a polynomial in $(Y_1,Y_2)$ which must vanish. This happens if and only if
\bea\label{n1234Contsr}
n_4=\frac{n_2 n_3}{n_1}\quad\mbox{and}\quad n_2=\pm n_1\,.
\eea
This leads to two branches for the solution (\ref{YsolnStart}).
\item Focusing on the $n_1=n_2$ branch, and requiring the coefficient of (\ref{HelpmSO5}) in front of $X_2$ to vanish, we find a first order equation for the function $g$:
\bea
(1+Y_2^2)\d_{Y_2} g+(1+Y_1^2)\d_{Y_1} g=0.
\eea
This reduces $g[Y_1,Y_2]$ to a function of one variable:
\bea\label{ODEbrn1}
n_2=n_1\quad\Rightarrow\quad
g=f\left[\frac{Y_1-Y_2}{1+Y_1Y_2}\right]\,.
\eea
Similarly, the $n_2=-n_1$ branch gives
\bea\label{ODEbrn2}
n_2=-n_1\ \Rightarrow\ (1+Y_1^2)\d_{Y_1} g-(1+Y_2^2)\d_{Y_2} g=0 \ \Rightarrow\
g=f\left[\frac{Y_1+Y_2}{1-Y_1Y_2}\right]
\eea
\item Substitution of (\ref{ODEbrn1}) or (\ref{ODEbrn2}) into (\ref{HelpmSO5}) reduces the Helmholtz equation  to a single ODE for the unknown function $f$, and the resulting normalizable wavefunctions $\Phi$ are given by
(\ref{YdepSO5a}).
\end{enumerate}
After this summary of the derivation we present the final result. The two branches of the solution (\ref{YsolnStart}) can be written as
\bea\label{YdepSO5a}
\Phi
&=&\frac{e^{2i[n_1(\alpha_L+\beta_L)+n_3(\alpha_R+\beta_R)]}}{(1+R^2)^q}
\frac{(1+y_-^2)^{{q+1}}}{y_-^{n_1-n_3}}
F\left[q+1-n_1,q+1+n_3;1-n_1+n_3;-y_-^2\right]\nn
\\
\Phi&=&\frac{e^{2i[n_1(\alpha_L-\beta_L)+n_3(\alpha_R-\beta_R)]}}{(1+R^2)^q}
\frac{(1+y_+^2)^{{q+1}}}{y_+^{n_1-n_3}}
F\left[q+1-n_1,q+1+n_3;1-n_1+n_3;-y_+^2\right],\nonumber
\eea
and complex conjugates of these expressions. Here we used the convenient variables $y_\pm$ 
defined in (\ref{SO5denom}):
\bea
y_\pm=\frac{Y_1\pm Y_2}{1\mp Y_1Y_2}\,.
\eea
Interestingly, the eigenvalues corresponding to functions (\ref{YdepSO5a})  depends only on $q$:
\bea\label{LamForYdep}
\Lambda=2q(2+q),
\eea
and parameters $(n_1,n_3)$ enter only through the constraints
\bea
\mbox{max}(n_1,n_3)\le q,\quad \mbox{integer}\quad (n_1,n_2,q).
\eea
Solutions (\ref{YdepSO5a}) can be dressed with functions of $R$ according to 
(\ref{SolnDressed})--(\ref{SolnDressedLam}):
\bea
&&\frac{1}{(1+R^2)^q}\rightarrow\frac{1}{(1+R^2)^q}F[-k,k+2q+3;2+2q;\frac{1}{1+R^2}],\nn
&&\Lambda=2q(2+q)+(2k+2q+3)^2-(2q+3)^2.
\eea
In two special cases, $n_1=n_3=q$ and $n_1=-n_3=-q$, the $y$--dependent parts of (\ref{YdepSO5a}) simplify to 
For $n_1=n_3=\nu$, the $z$--dependent parts simplify to
\bea
\frac{1}{(1+y_\pm^2)^{q}}=\left[\frac{(1\pm Y_1Y_2)^2}{(1+Y_1^2)(1+Y_2^2)}\right]^{q}\quad
\mbox{and}\quad
\left[\frac{y_\pm^2}{1+y_\pm^2}\right]^{q}=\left[\frac{(Y_1\pm Y_2)^2}{(1+Y_1^2)(1+Y_2^2)}\right]^{q}\,,\nonumber
\eea
leading to pure powers of the expressions (\ref{YsolnEx}). 

To summarize, we have demonstrated that the $X$--independent ansatz (\ref{YsolnStart}) introduces constraints (\ref{n1234Contsr}) on the $[U(1)]^4$ charges and reduces $g[Y_1,Y_2]$ to a function of one variable. This implies, that the solution (\ref{YsolnStart}) depends on three parameters: two combinations of 
$(n_1,n_2,n_3,n_4)$ which are not eliminated by the constraint (\ref{n1234Contsr}), and an additional integer coming from the solutions of the ODE for the function $g$. Dressing the solutions (\ref{YdepSO5a}) with a function of $R$ introduces the fourth parameter. Since the most general solution of the Helmholtz equation  (\ref{HelpmSO5}) is expected to depend on $10$ parameters, clearly the wavefunctions (\ref{YsolnStart}) form a very small subset. Unfortunately, the nice separability encountered in (\ref{YsolnStart}) does not persist for the $X$--dependent functions, but several infinite families of wavefunctions can be constructed, and they will be discussed in the next subsection. 

\subsection{Solutions linear in $X$ coordinates}
\label{SecSubXlinear}

In the previous subsection we have constructed the most general $X$--independent solution of the Helmholtz equation  (\ref{HelpmSO5}). Unfortunately, explicit closed--form expressions for all $X$ dependent eigenfunctions are unlikely to exist\footnote{Procedure (\ref{PhiYoung}) allows to construct all such functions algorithmically, but the combinatorics becomes complicated.}. Nevertheless in this subsection we construct several infinite families of $X$--dependent eigenfunctions, and these results can be viewed as a complement of the algebraic procedure (\ref{PhiYoung}), which is practical only for representations with a small number of boxes in the Young diagrams. 

\bigskip

Let us look at wavefunctions which are linear in $(X_1,X_2,X_3,X_4)$. The explicit examples (\ref{eqn126}) suggest that it might be useful to write the solutions in terms of complex variables 
$(z_1,{\bar z}_1,z_2,{\bar z}_2)$. Let us impose an ansatz
\bea\label{LinearZform}
\Phi=\frac{e^{2i[n_1\alpha_L+n_2\beta_L+n_3\alpha_R+n_4\beta_R]}}{(1+R^2)^q}\Big[z_1 g_1(Y_1,Y_2)+\bar{z}_1 g_2 (Y_1,Y_2)+z_2 g_3(Y_1,Y_2)+\bar{z}_2 g_4(Y_1,Y_2)\Big]\nn
\eea 
Substituting this function into the equation (\ref{HelpmSO5}), and requiring the coefficients in front of eight combinations $(X_1^2 X_2,X_1^3,X_2^2 X_1,X_2^3,X_3^2 X_4,X_3^3,X_4^2 X_3,X_4^3)$, to vanish, we can {\it algebraically} solve the resulting {equations} for the eight second derivatives
\bea 
\p^2_{Y_1} g_i(Y_1,Y_2),\quad \p^2_{Y_2} g_i(Y_1,Y_2).
\eea 
Substituting the result back to (\ref{HelpmSO5}), we observe that the coefficients in front of ($X_1^2 X_3$,
\\ $X_2^2 X_3$,\,$X_3^2 X_1$,\,$X_4^2 X_2$) contain only functions $g_i$, but not  their derivatives. Requirement of  having non-trivial solutions implies that the determinant of the characteristic matrix has to vanish. This condition leads to only eight possibilities:
\bea\label{tempDec24}
&&n_2=n_1\pm 1, \quad n_3=n_4;\nn
&&n_2=-n_1\pm 1,\quad n_3=-n_4;\\
&&n_4=n_3\pm 1, \quad n_1=n_2;\nn
&&n_4=-n_3\pm 1,\quad n_1=-n_2.\nonumber
\eea
Some of the resulting solutions can be obtained from the others by applying discrete symmetries of the metric. First, by taking a complex conjugate of the solution, if necessary, we can focus only on ``$-1$'' option instead of $\pm 1$. Furthermore, the first two options in (\ref{tempDec24}), are related by changing the signs of $(\alpha_L,\alpha_R)$. While such change by itself is not a symmetry of the metric, it is a part of a larger one:
\bea\label{SymmMap49}
(\alpha_L,\alpha_R,Y_1,X_1)\rightarrow -(\alpha_L,\alpha_R,Y_1,X_1).
\eea
This symmetry also interchanges the last two options in (\ref{tempDec24}). Therefore, there are two genuinely 
distinct possibilities:
\bea
n_2=n_1-1, \quad n_3=n_4\quad \mbox{and}\quad n_4=n_3- 1, \quad n_1=n_2\,.
\eea
If one of these constraints is imposed, some of the algebraic equations for $g_i$ can be solved, and the results are
\bea\label{LinearZbranches}
n_2=n_1-1,\ n_3=n_4:&&\hskip -0.5cm
\Psi=\frac{E_{n_1,n_2,n_3,n_4}}{(1+R^2)^q}\left[g_1(Y_1,Y_2){\bar z}_1+g_2(Y_1,Y_2){z}_2\right],\nn
\\
n_1=n_2,\ n_4=n_3-1:&&\hskip -0.5cm
\Psi=\frac{E_{n_1,n_2,n_3,n_4}}{(1+R^2)^q}\left[\left\{{\bar z}_2+\frac{Y_1+Y_2}{1-Y_1Y_2}{\bar z}_1\right\}g_4+\left\{{z}_2+\frac{Y_1Y_2-1}{Y_1+Y_2} z_1\right\}g_3\right].\nonumber
\eea
The differential equations for these two ansatze are analyzed in the Appendix \ref{AppSO5Linear}, and they lead to the following solutions.
\begin{itemize}
\item
For the first option in (\ref{LinearZbranches}), functions $g_1$ and $g_2$ can depend on the $(Y_1,Y_2)$ coordinates only through the combination $y_-$ defined in (\ref{SO5denom}):
\bea
g_{1}[Y_1,Y_2]=h_{1}\left[y_-\right],\quad g_{2}[Y_1,Y_2]=h_{2}\left[y_-\right],\quad 
 y_-=\frac{Y_1-Y_2}{1+Y_1 Y_2}\,.
\eea
Function $h_1$ and $h_2$ satisfy an overdetermined system of ordinary differential equations, and one of the consistency conditions implies that 
\bea\label{h1Branch1}
h_1(w)=\frac{1}{2w\sigma}\left[-w(1+w^2)h_2'+[n_3(1+w^2)-n_1(1-w^2)]h_2\right]
\eea
with some constant $\sigma$. The remaining equations lead to the expressions for $(\Lambda,\sigma)$ in terms of the parameters $(q,n_1,n_3)$ of the ansatz (\ref{LinearZbranches}), and all regular solutions can be divided into two branches:
\bea\label{LinearZsolnE1}
&&\hskip -0.9cm\mbox{(a)}:\quad \Lambda=2q(q+2),\ \sigma=n_1+q,\nn
&&\hskip -0.5cm n_1> n_3-1\ \Rightarrow\ h_2[w]=\frac{w^{n_1-n_3}}{(1+w^2)^q}F[n_1-q,-n_3-q;n_1+1-n_3;-w^2],
\nn
&&\hskip -0.5cm n_3> n_1-1\ \Rightarrow\ h_2[w]=\frac{w^{n_3-n_1}}{(1+w^2)^q}F[n_3-q,-n_1-q;n_3+1-n_1;-w^2];
\nn
&&\hskip -0.9cm\mbox{(b)}:\quad  \Lambda=2q(q+1),\ \sigma=n_1-q, \\
&&\hskip -0.5cm
n_1> n_3-1\ \Rightarrow\ h_2[w]=\frac{w^{n_1-n_3}}{(1+w^2)^{q-1}}F[1+n_1-q,1-n_3-q;n_1+1-n_3;-w^2],
\nn
&&\hskip -0.5cm
n_3> n_1-1\ \Rightarrow\ h_2[w]=\frac{w^{n_3-n_1}}{(1+w^2)^{q-1}}F[1+n_3-q,1-n_1-q;n_3+1-n_1;-w^2].
\nonumber
\eea
As in the $X$--independent case (\ref{YdepSO5a}), (\ref{LamForYdep}), the eigenvalue $\Lambda$ depends only on 
$q$, and the $[U(1)]^4$ charges enter only through the regularity bounds: the first arguments of the hypergeometric functions appearing in (\ref{LinearZsolnE1}) must be non--positive integers. 
\item
For the second option in (\ref{LinearZbranches}), functions $g_3$ and $g_4$ can depend on the $(Y_1,Y_2)$ coordinates only through the combination $y_-$ defined in (\ref{SO5denom}), up to a fixed prefactor. Specifically, the second line in (\ref{LinearZbranches}) must have the form
\bea\label{LongBranchStart}
&&\hskip -1.2cm n_1=n_2,\ n_4=n_3-1:\\
&&\hskip -1.2cm
\Psi=\frac{E_{n_1,n_2,n_3,n_4}}{(1+R^2)^q}\frac{Y_1+Y_2}{\sqrt{1+Y_2^2}\sqrt{1+Y_1^2}}\left[
\left\{{\bar z}_1+\frac{1}{y_+}{\bar z}_2\right\}{\hat g}_4[y_-]+\left\{{z}_2-\frac{1}{y_+} z_1\right\}{\hat g}_3[y_-]\right]\nonumber
\eea
Functions ${\hat g}_3$ and ${\hat g}_4$ satisfy an overdetermined system of ODEs, and one of the consistency conditions gives
\bea\label{LongBranch2}
{\hat g}_3=-y {\hat g}_4+h,\quad 
{\hat g}_4=\frac{[1 + 2(\Lambda-3q- 2q^2)y^2 + (n_1-n_3) (1 + y^2)] h + 
 y (1 + y^2) h'}{2 (1 - n_3 +\Lambda - 3q - 
   2q^2) y (1 + y^2)}\nonumber\\
\eea
The remaining equations lead to two branches for function $h$, the counterparts 
of (\ref{LinearZsolnE1}) for the present case:
\bea\label{LinearZsolnE2}
&&\hskip -0.9cm\mbox{(a)}:\quad \Lambda=2q(q+2),\nn
&&\hskip -0.5cm n_1> n_3-2\ \Rightarrow\ h[w]=\frac{w^{n_1-n_3+1}}{(1+w^2)^{q-\frac{1}{2}}}F[n_1-q,1-n_3-q;2+n_1-n_3;-w^2],
\nn
&&\hskip -0.5cm n_3> n_1\ \Rightarrow\ h[w]=\frac{w^{n_3-n_1-1}}{(1+w^2)^{q-\frac{1}{2}}}
F[n_3-q-1,-n_1-q;n_3-n_1;-w^2];
\nn
&&\hskip -0.9cm\mbox{(b)}:\quad  \Lambda=2q(q+1), \\
&&\hskip -0.5cm
n_1> n_3-2\ \Rightarrow\ h[w]=\frac{w^{1+n_1-n_3}}{(1+w^2)^{q-\frac{3}{2}}}F[1+n_1-q,2-n_3-q;2+n_1-n_3;-w^2],
\nn
&&\hskip -0.5cm
n_3> n_1\ \Rightarrow\ h[w]=\frac{w^{n_3-n_1-1}}{(1+w^2)^{q-\frac{3}{2}}}F[n_3-q,1-n_1-q;n_3-n_1;-w^2].
\nonumber
\eea
Once again, the eigenvalue $\Lambda$ depends only on 
$q$, and the $[U(1)]^4$ charges enter only through the regularity bounds: the first arguments of the hypergeometric functions appearing in (\ref{LinearZsolnE2}) must be non--positive integers. 
\end{itemize}
To summarize, we have constructed all wavefunctions which have the form (\ref{LinearZform}). Apart from the $R$--dependence these solutions are linear in $(X_1,X_2,X_3,X_4)$ coordinates. We have shown that there are only eight possibilities (\ref{tempDec24}) for the $[U(1)]^4$ charges, and for each of these options the final answers are specified by three quantum numbers $(n_1,n_3,q)$. The solutions can be divided into several groups:
\begin{enumerate}[(i)]
\item For $n_2=n_1-1$, $n_4=n_3$, the wavefunctions are given by 
\bea\label{tempSolnBrn1}
\Psi=\frac{E_{n_1,n_1-1,n_3,n_3}}{(1+R^2)^q}\Big[h_1[y_-]{\bar z}_1+h_2[y_-]{z}_2\Big]
\eea
with functions $h_1$ and $h_2$ given by (\ref{h1Branch1}) and (\ref{LinearZsolnE1}). 
\item The wavefunctions with $(n_2,n_4)=(-n_1-1,-n_3)$ are obtained by applying the map (\ref{SymmMap49}) to the solution (\ref{tempSolnBrn1}). The result reads
\bea\label{tempSolnBrn1e2}
\Psi=\frac{E_{-n_1,n_1-1,-n_3,n_3}}{(1+R^2)^q}\Big[-h_1[-y_+]{z}_1+h_2[-y_+]{z}_2\Big]\,.
\eea
\item The solutions with $(n_2,n_4)=(\pm n_1+1,\pm n_3)$ are constructed by taking complex conjugates of (\ref{tempSolnBrn1}) and (\ref{tempSolnBrn1e2}). The four branches covered by the items (i)--(iii) have the general structure similar to (\ref{tempSolnBrn1}). 
\item The solution with $(n_2,n_4)=(n_1,n_3-1)$ has the form (\ref{LongBranchStart}) with various functions given by (\ref{LongBranch2}) and (\ref{LinearZsolnE2}). 
\item The wavefunctions with  $(n_2,n_4)=(-n_1,-n_3-1)$ are obtained by applying the transformation (\ref{SymmMap49}) to equations (\ref{LongBranchStart}), (\ref{LongBranch2}), and (\ref{LinearZsolnE2}).
\item The solutions with  $(n_2,n_4)=(\pm n_1,\pm n_3+1)$ are obtained by taking complex conjugates of the wavefunctions from the items (iv) and (v). All four solutions covered by the items (iv)--(vi) have the general structure of the state (\ref{LongBranchStart}).
\end{enumerate}
Each of the eight branches discussed in this subsection has wavefunctions which are specified by three integer parameters $(n_1,n_3,q)$. The fourth quantum can be added by dressing the solutions using (\ref{SolnDressed}) and (\ref{SolnDressedLam}) with $p=1$. Therefore, we have constructed eight four--parameter branches of scalar wavefunctions with linear dependence on individual $X_i$ and a complicated dependence on $R$. 

\subsection{Separable states in symmetric representations}
\label{SecSubSymRep}

In the last few subsections we have constructed several infinite families of scalar eigenfunctions by solving the differential equation (\ref{HelpmSO5}). Alternatively, one can use the algebraic method (\ref{PhiYoung}), but unfortunately it involves combinatorics which becomes very complicated as the size of a representation grows. Nevertheless, the construction (\ref{PhiYoung}) can be used to find some infinite families of wavefunctions in a closed form, and in this subsection we will do so for the fully symmetric representations of $SO(5)$. We will begin with re-casing the family (\ref{NeutralHyper}) as a summation (\ref{PhiYoung}) for the symmetric representations, and then we will extend this construction to more general wavefunctions in such representation. 

While it is very easy to find the family (\ref{NeutralHyper}) using differential equations (one just needs to use the general structure (\ref{Rsplit}) of the metric), it is instructive to recover these solutions from the group theoretic construction (\ref{PhiYoung}). Since solutions (\ref{NeutralHyper}) are neutral under $[U(1)]^2$, and they depend only on $R$, but not on the individual coordinates $(X_1,X_2,X_3,X_4,Y_1,Y_2)$, it is clear that the wavefunctions are built only from $g_{55}$ in the parameterization (\ref{SO5Param}). With only one available matrix element, the wavefunction (\ref{PhiYoung}) vanishes unless the representation is fully symmetric, and in the latter case one has
\bea\label{Phi55}
\Phi=g_{55}\dots g_{55}-(traces).
\eea
After recovering the family (\ref{NeutralHyper}) from the last equation, we will analyze more general states constructed from products of $g_{a5}$: once again, since all ingredients have the same second index, only symmetric representations are allowed, and the expression (\ref{PhiYoung}) reduces to
\bea\label{Phi5a}
\Phi=g_{a_15}\dots g_{a_k 5}-(traces).
\eea 
The structure of traces in (\ref{Phi55}) and (\ref{Phi5a}) will be specified below. 

We begin with analyzing the wavefunction (\ref{Phi55}). In general, the state in a symmetric representation is\footnote{We are focusing on the even number of boxes ($2s$) in the Young tableau, and the odd case can be discussed in the same way.}
\bea\label{tempDec25}
\Phi=g_{a_1 b_1}\dots g_{a_{2s} b_{2s}}+\sum_{k=1}^s(-1)^k\frac{(4s+1-2k)!!}{(4s+1)!!}\delta_k\Pi_{2s-2k}\,,
\eea
where we defined a shorthand notation
\bea
\delta_k\Pi_{2s-2k}\equiv \sum_{i_1\dots i_{2k}}\left\{\left[\delta_{a_{i_1}a_{i_2}}\dots 
\delta_{a_{i_{2k-1}}a_{i_{2k}}}\delta_{b_{i_1}b_{i_2}}\dots 
\delta_{b_{i_{2k-1}}b_{i_{2k}}}\right]\left[\frac{g_{a_1 b_1}\dots g_{a_{2s} b_{2s}}}{
g_{a_{i_1}b_{i_1}}\dots g_{a_{i_{2k}}b_{i_{2k}}}}\right]\right\}
\eea
Combinatorial factors in (\ref{tempDec25}) are determined by requiring the contaction with respect to any pair of indices $(a_i,a_j)$ to vanish. For example, observing that\footnote{Recall that since $g$ is an element of $SO(5)$, it satisfies the orthogonality relation $gg^T=1$.} 
\bea
\delta^{a_1a_2}g_{a_1 b_1}\dots g_{a_{2s} b_{2s}}=\delta_{b_1b_2}g_{a_3 b_3}\dots g_{a_{2s} b_{2s}}
\nonumber
\eea
and
\bea
\delta^{a_1a_2}\delta_1\Pi_{2s-2}=(5+2(2s-2))\delta_{b_1b_2}g_{a_3 b_3}\dots g_{a_{2s} b_{2s}}+\mbox{(terms with fewer $g$)},
\nonumber
\eea
we conclude that the coefficient in front of the $k=1$ term in (\ref{tempDec25}) is indeed 
$-\frac{1}{4s+1}$. The other coefficients are determined using induction. 

To apply equation (\ref{tempDec25}) to the state (\ref{Phi55}), we observe that if all indices $a_i=b_j=5$, then 
\bea\label{tempDec25q}
\delta_k\Pi_{2s-2k}=\frac{(2s)!}{(2s-2k)! s!}(g_{55})^{2s-2k}\,.
\eea
Substitution into (\ref{tempDec25}) gives
\bea\label{tempDec25a}
\Phi=(g_{55})^{2s}+\sum_{k=1}^s(-1)^k\frac{(4s+1-2k)!!}{(4s+1)!!}\frac{(2s)!}{(2s-2k)! s!}(g_{55})^{2s-2k}
\propto F[-s,\frac{3}{2}+s;\frac{1}{2};(g_{55})^2].\nn
\eea
To relate this answer to the solution (\ref{NeutralHyper}) we observe that equation (\ref{SO5Param}) gives
\bea
g_{55}=\frac{1-R^2}{1+R^2}\,,\nonumber
\eea
and that function (\ref{NeutralHyper}) can be rewritten as
\bea\label{tempDec25b}
\Phi=c_1 F\left[-\frac{k}{2},\frac{3+k}{2};\frac{1}{2};\left(g_{55}\right)^2\right]+
c_2 g_{55}F\left[-\frac{k-1}{2},\frac{4+k}{2};\frac{3}{2};\left(g_{55}\right)^2\right]\,.
\eea
The numerical coefficients $(c_1,c_2)$ are such that $c_1=0$ for the odd values of $k$, and $c_2=0$ for the even ones. Clearly, there is a perfect agreement between (\ref{tempDec25a}) and (\ref{tempDec25b}) for $k=2s$, and the case of odd $k$ can be analyzed in the same way\footnote{One would have to start with a counterpart of equation (\ref{tempDec25}) for the representations with an odd number of boxes in the Young tableaux.}. Therefore, the algebraic construction (\ref{PhiYoung}) reproduces the family (\ref{NeutralHyper}). We went though this derivation to illustrate the procedure for analyzing (\ref{tempDec25}) in a simple setting, and now we will present the results for more complicated cases. The derivation follows the same logic, but the technical details are more involved, and they are presented in the Appendix \ref{AppSO5comb}. 

Let us go back to the states in the symmetric representation (\ref{tempDec25}), take all indices $b_k$ to be equal to five, and allow indices $a_k$ to take values from one to four. In other words, we are looking at states (\ref{Phi5a}). The experience gained in the previous subsections, suggests that it is convenient to use complex coordinates $z_1=X_1+iX_2$ and $z_2=X_3+iX_4$ instead of $X_j$, and to accommodate this change of coordinates we define\footnote{We use $(z,w)$ instead of $(z_1,z_2)$ to avoid double superscript in various expressions written below. Also, our results will be applicable to solutions which depend on different $(z,w)$ complex structures, for example for $z=X_1-iX_3$, $w=X_2-iX_4$, however, such combinations will not carry specific charges under $[U(1)]^4$ symmetries corresponding to translations in $(\alpha_L,\beta_L,\alpha_R,\beta_R)$.}
\bea\label{ComplexGcomb}
g_{z5}=g_{15}+ig_{25},\quad g_{w5}=g_{35}+ig_{45}\,.
\eea
To perform contractions of these ingredients one can use the relevant Kronecker symbols
\bea\label{CombinKron}
\delta_{zz}=\delta_{x_1+i x_2,x_1+i x_2}=0,\quad \delta_{z{\bar z}}=\delta_{x_1+i x_2,x_1-i x_2}=2,\quad
\delta_{ww}=0,\quad \delta_{w{\bar w}}=2.
\eea
For example, let us consider a wavefunction built only from $g_{z5}$ and $g_{{\bar z}5}$. Let us assign charge one to $g_{z5}$ and charge minus one $g_{{\bar z}5}$. It is clear that a contraction cannot change the charge of the product, so all terms in (\ref{tempDec25}) have the same. Assuming that this charge $q$ is non--negative, we conclude that the last term in (\ref{tempDec25}) is equal to $(g_{z5})^q$ multiplied by a constant. For other terms one finds a counterpart of (\ref{tempDec25q}):
\bea\label{tempDec25qa}
\delta_k\Pi_{q+2s-2k}=2^k k!\left[\frac{s!}{k!(s-k)!}\right]\left[\frac{(s+q)!}{k!(s+q-k)!}\right]
(g_{z5})^{q+s-k}(g_{{\bar z}5})^{s-k}\,.
\eea
Then the sum (\ref{tempDec25}) can be easily performed, and it gives
\bea\label{eqn146t}
\Phi=(g_{z5})^q F[-k,\frac{3}{2}+k+q;1+q;(g_{z5}g_{\bar z 5})]\,.
\eea
The solution with a negative $q$ is obtained by a making a replacement 
$g_{z5}\leftrightarrow g_{\bar z 5}$. 

Once all $g_{a5}$ are included, the combinatorics becomes more complicated, but the final result is rather compact:
\bea\label{FinalProdGeom}
&&\hskip -1.2cm
\Phi=(g_{z5})^{q_1}(g_{w5})^{q_2}
F_2\left[\frac{3}{2}+q_1+q_2+k_1+k_2,-k_1,-k_2;q_1+1,q_2+1;g_{z5}g_{\bar z 5},g_{w5}g_{\bar w 5}
\right],\nn
&&\hskip -1.2cm \Lambda=(2k_1+2k_2+q_1+q_2)(3+2k_1+2k_2+q_1+q_2).
\eea
Here $F_2$ is the Appell's generalizion of the hypergeometric function defined by the series expansion\footnote{Recall the standard notation $(a)_n=a\dots (a+n-1)$.}
\bea\label{ApellDef}
F_2[a,b_1,b_2;c_1,c_2;x,y]=\sum_{m,n=0}^\infty \frac{(a)_{m+n}(b_1)_m(b_2)_n}{
m!n! (c_1)_m(c_2)_n}x^my^n\,.
\eea
Regularity requires $(k_1,k_2)$ to be non--negative integers.
Combinatorial derivation of the expression (\ref{FinalProdGeom}) is presented in the Appendix \ref{AppSO5comb}.

Expression (\ref{FinalProdGeom}) reduces to the standard hypergeometric function in the special case (\ref{eqn146t}) and its simple extension
\bea\label{eqn146extra}
\Phi=
(g_{z5})^{q_1}(g_{w5})^{q_2}F[-k,\frac{3}{2}+k+q_1+q_2;1+q_1;g_{z5}g_{\bar z 5}],\quad s=q_1+q_2+2k.
\eea
with various permutations of indices $(z,{\bar z},w,{\bar w})$. Note that solutions (\ref{FinalProdGeom}) and (\ref{eqn146extra}) carry specific $[U(1)]^4$ charges, but their $R$--dependence is rather complicated. This can be seen from the explicit form of the matrix elements:
\bea
g_{z5}=-\frac{2e^{-2i\alpha_L}z}{1+R^2},\quad g_{w5}=-\frac{2e^{-2i\beta_L}w}{1+R^2}\,.
\eea
In particular, the solutions described in this subsection don't have the form (\ref{SolnToDress}), so they cannot be dressed with additional functions of $R$ using the procedure described in section \ref{SecDressR}. Furthermore, all arguments of this subsection are equally applicable to $g_{5a}$, so by making replacements
\bea\label{SO5gFlip}
g_{a5}\rightarrow g_{5a}
\eea
in the expressions (\ref{FinalProdGeom}), (\ref{eqn146extra}), one still gets solutions of the Helmholtz equation (\ref{HelpmSO5}) with the same eigenvalues. The resulting separation is perhaps even more impressive since the relevant matrix elements are more complicated
\bea
g_{z5}=\frac{2e^{-2i\alpha_R}(Z_{1-}-y_+Z_{2-})}{D}\,,\quad
g_{w5}=\frac{2e^{-2i\beta_R}(Z_{2-}+y_+Z_{1-})}{D}\,.
\eea
We used the convenient notation introduced in (\ref{SO5denom})--(\ref{SO5ComplCord}). 

We conclude this subsection by counting parameters. There are $12$ branches of (\ref{FinalProdGeom}) corresponding to different choices of complex structures\footnote{Here we focus only on the complex structuctures in $(X_1,X_2,X_2,X_4)$ space. There are also wavefunctions obtained from (\ref{ComplexGcomb}) and (\ref{FinalProdGeom}) by replacements like $X_1\leftrightarrow X_5$, but they have more complicated coordinate dependence, so we have not studied them in detail.}, and $12$ more branches of the solutions flipped by
(\ref{SO5gFlip}). Each branch has four integer parameters $(q_1,q_2,k_1,k_2)$. Interestingly, each of the solutions (\ref{YdepSO5a}) dressed with functions of $R$ had four parameters as well. It would be very interesting to find separable solutions with a larger number of free parameters since a general solution in $10$ dimensions should be parameterized by ten numbers.

\subsection{Separation in terms of spherical harmonics}
\label{SecSubSpehere}

To conclude the discussion of separable solutions in the $SO(5)$ WZW model, we also mention 
an alternative parameterization of the group element (\ref{SO5Param}) that leads to another set of eigenfunctions depending on four coordinates. This alternative parameterization has a major disadvantage in comparison to (\ref{SO5Param}): the Cartan group of $SO(5)\times SO(5)$ acts in a complicated way\footnote{Recall that the WZW model on a group $G$ has $G\times G$ global symmetries acting by $g\rightarrow h_L g h_R$.}, so we are discussing it only for completeness.  

Let is consider a scalar field on the background of an $SO(N)$ WZW model. A general group element $g$ can be written as 
\bea 
g=q\,h,
\eea 
where $q$ is an element of the subgroup $SO(N-1)$ and $h$ is a coset representative of $SO(N)/SO(N-1)$ which describes the symmetric space $S^{N-1}$. Then the left-invariant one-form frames are given by
\bea 
\mathcal{L}=g^{-1}dg=h^{-1}(dh h^{-1}+q^{-1}dq)h\equiv h^{-1}{\tilde{\mathcal{L}}}h.
\eea 
The action (\ref{WZWaction}), can be rewritten in terms of ${\tilde{\mathcal{L}}}$,
\bea
S=-\frac{k}{2\pi}\int d^2\sigma \eta^{\alpha\beta}\mbox{tr}(
{\tilde{\mathcal{L}}}_\alpha {\tilde{\mathcal{L}}}_\beta )+
\frac{ik}{6\pi}\int \mbox{tr}({\tilde{\mathcal{L}}}\wedge  {\tilde{\mathcal{L}}}\wedge  
{\tilde{\mathcal{L}}})\,,\nonumber
\eea
so these objects can be used as frames. Introducing a split between the group and the coset,
\bea 
{\tilde{\mathcal{L}}}=dh h^{-1}+q^{-1}dq\equiv R_h+L_q
\eea 
we can write the frames as a lower triangular block matrix:
\bea 
{\tilde{\mathcal{L}}}^{\bf A}_{~M}=\begin{bmatrix}
R^{\alpha}_{~\mu}&0\\
R^{a}_{~\mu}& L^a_{~m}
\end{bmatrix},\quad A=(\alpha,a),\quad M=(\mu,m).
\eea 
where we have used the Greek (Latin) indexes are denoting the coset (subgroup) projections.
The inverse of this  lower triangular block matrix leads to the frames with contravariant indices\footnote{Note that $R^{a}_{~\nu} R^{~\nu}_{\alpha} \neq \delta^a_\alpha$} 
\bea
\hat{e}^{M}_{~\bf A}=\begin{bmatrix}
R^{\mu}_{~\alpha}&0\\
-L^{m}_{~a}R^{a}_{~\nu} R^{\nu}_{~\alpha}&  L^{m}_{~\,a}
\end{bmatrix}\equiv \begin{bmatrix}
R^{\mu}_{~\alpha}&0\\
N^m_{~\,\alpha}& L^{m}_{~\,a}
\end{bmatrix},
\eea 
and the resulting inverse metric has a fibered structure:
\bea 
G^{MN}=\hat{e}_{\bf A}^{~M} \hat{e}_{\bf B}^{~N} \eta^{\bf AB}=\eta^{\alpha\beta}(R_\alpha^{~\mu}+N_\alpha^{~m})(R_\beta^{~\nu}+N_\beta^{~m})+\eta^{ab}L_a^{~m}L_b^{~n}\,.
\eea 
In particular, if we look at the Helmholtz equation (\ref{HelpmSO5}) and assume that the scalar $\Phi$ depends only on the coset coordinates, then the problem reduces to the eigenvalue equation on the sphere $S^{N-1}$. The eigenvalies are given by
\bea
\Lambda=k(k+N-2),
\eea
and eigenfunctions are fully separable an several coordinate systems. For example, writing the metric of $S^{p}$ as 
\bea
ds^2_p=d\theta_p^2+\sin^2\theta_p ds^2_{p-1},
\eea
we can build the eigenfunctions using induction:
\bea
\Phi=\prod_p^{N-1} \Phi_p(\theta_p).
\eea
Function $\Phi_p(\theta_p)$ satisfies an ordinary differential equation
\bea
\frac{1}{(\sin\theta_p)^{p-1}}
\frac{d}{d\theta_p}\left[(\sin\theta_p)^{p-1}\frac{d\Phi_p}{d\theta_p}\right]-\frac{\Lambda_{p-1}\Phi_p}{\sin^2\theta_p}+\Lambda_p\Phi_p=0\,.
\eea
The normalizable solutions of this equation can be written in terms of the associated Legendre polynomials:
\bea
\Phi_p=(\sin\theta_p)^{\frac{2-p}{2}} P^{(\mu)}_\la(\cos\theta_p),\quad
\la=k_p-1+\frac{p}{2},\quad \mu=k_{p-1}-1+\frac{p}{2}\,.
\eea
In the case of the $S^{N-1}$ dimensional sphere, the solution depends on $(N-1)$ integer parameters $(k_1,\dots k_{N_1})$. Interestingly, for $SO(5)$ we find four parameters, the same number that has been encountered elsewhere in this section, although the spherical separation is very different form the other constructions discussed here. In the $SO(4)$ case spherical separation gives only three parameters, in contrast to the six--parameter separation encountered in section \ref{SecSubSO4}. It would be interesting to see whether for larger groups spherical separation becomes more or less powerful than the one coming from parameterizations like (\ref{WZWSO5}). 

\section{Discussion}

In this article we have studied equations for scalar and vector fields on backgrounds of several (gauged) WZW models. While the scalar spectrum has been known for some time \cite{PS}, the algebraic construction of the relevant eigenfunctions turns our to be rather involved \cite{LTian}, and it is desirable to look for a more explicit form of the solutions. Furthermore, the CFT construction of eigenvalues and eigenfunctions \cite{PS,LTian} does not seem to be easily extendable to vector and tensor fields. To cure these problems, we focused on extracting the wavefunctions directly from the perspective of field equations instead of appealing to algebraic methods. 

For the $SO(4)$ group and its cosets, we demonstrated the full separation of variables in equations for the scalar and vector fields, including the vectors with field strength twisted by the $B$--field. Although we derived separation of variables from the first principles, therefore establishing uniqueness of the separable ansatz for the vector components, we found that the final expressions have the same structure as their counterparts for the rotating black holes \cite{LMaxw}. This suggests universality of the form (\ref{AnstzForSU2a}) for the separable components of vector fields on all geometries that admit separation, and it would be very interesting to test this hypothesis on other backgrounds.

For the $SO(5)$ group, the full separation of scalar and vector equations seems unlikely, but we found several classes of separable solutions. All our families are parameterized by four integers, in contrast to ten quantum numbers expected for the ten--dimensional geometry. It would be interesting to either find separable families with more parameters or to understand why this can't be done. It would also be interesting to study scalar and vector fields on manifolds corresponding to larger groups and cosets.

\section*{Acknowledgements}

This work was supported in part by the DOE grant DE-SC0017962 (OL) and by the UCAS program of the Special Research Associate, as well as the internal funds of the KITS (JT).

\appendix

\section{Vector field on the $SU(2)$ WZW model}
\label{AppSU2}
\renewcommand{\theequation}{A.\arabic{equation}}
\setcounter{equation}{0}
In section \ref{SecSubSub2} we outlined the procedure for finding vectors modes on $SU(2)$, and in this appendix we present the technical details of this derivation. We will perform the analysis in two steps. In section \ref{AppSU2sub1} we will focus on the standard equations for the vector ((\ref{temp1Fij}) with 
$\zeta=0$) and give technical details supporting the steps (i)-(iv) presented in section \ref{SecSubSub2}. This justifies the ansatz (\ref{AnstzForSU2}) and prove its uniqueness. In section \ref{AppSU2sub1}, we will impose the ansatz (\ref{AnstzForSU2}) for the modified vector equation (\ref{temp1Fij}) and derive the relations (\ref{AvaluesZeta}) and (\ref{VecEigen}). This would justify step (v) in section \ref{SecSubSub2}.

\subsection{Standard equation for the vector: $\zeta=0$.}
\label{AppSU2sub1}

In this subsection we will focus on $\zeta=0$ case to justify the ansatz (\ref{AnstzForSU2}) and prove its uniqueness. To get some intuition about the structure of relevant components we will begin with a special case $n_1=n_2=\zeta=0$ of the system (\ref{temp1Fij}), (\ref{CfieldFormSU2}). Then we extend the analysis to arbitrary values of $(n_1,n_2)$. 

We begin with recalling that the most general separable solution for a vector field in the geometry (\ref{MetrSO4y1}) is given by
\bea
C_i dx_i=e^{i n_1\gamma_L+in_2\gamma_R}\left[V_y dy_1+V_1 d\gamma_L+V_2 d\gamma_R\right],
\eea
where $(V_y,V_1,V_2)$ are functions of $y_1$, which are mixed in equations  (\ref{temp1Fij}). We are looking for combinations of these components that satisfy decoupled equations, and to get insights into the structure of such combinations, we begin with studying a special case:
\bea\label{VecNoncycl}
n_1=n_2=0,\quad \zeta=0.
\eea
Once these conditions are imposed, the mode $V_y$ decouples, it gives $\lambda_{vector}=0$, while function $V_y$ remains arbitrary. This ``pure gauge'' describes the scalar mode (\ref{VectScalBranch}) in the special case (\ref{VecNoncycl}). To decouple the remaining components, we introduce linear combinations 
$V_\pm$:
\bea\label{tempVpm}
V_\pm=\frac{V_1\pm V_2}{2}:\quad C_i dx_i=
\left[V_y dy_1+V_+ (d\gamma_L+d\gamma_R)+V_- (d\gamma_L-d\gamma_R)\right].
\eea
Then the system (\ref{temp1Fij}) with $\lambda_{vector}\ne 0$ reduces to differential equations for 
$V_\pm$ 
\bea\label{DiffEqnApm}
( y_1\pm 1)\frac{d}{dy_1}\left[(y_1\mp 1)V_{\pm}'\right]+\lambda_{vector}V_{\pm}=0.
\eea
The normalizable solutions are
\bea
V_\pm(y_1)=C_\pm F\left[-M,M;1;\frac{1\mp y_1}{2}\right],\quad \lambda_{vector}=-M^2,
\eea
where $M$ is a non--negative integer. Comparing this with (\ref{B1hyper}) for $n_+=n_-=0$, we can identify $M$ with $L$ and express the vector modes in terms of the solutions of the scalar equation (\ref{SO4scalODE}):
\bea\label{VpmScalSpecial}
V_\pm(y_1)=C_\pm\left[(1-y_1^2)B_1'+M(y_1\pm 1)B_1\right]
\eea
Then equations (\ref{DiffEqnApm}) with $\lambda_{vector}=-M^2$ reduce to an ODE for $B_1$:  
\bea
\frac{d}{dy_1}\left[(y_1^2-1)B_1'\right]-M(M+1) {B_1}=0,
\eea
which is a special case of (\ref{SO4scalODE}).

\bigskip

Next we relax the values $(n_1,n_2)$ while keeping $\zeta=0$. In addition to the pure gauge, 
$C_i dx^i=dC(x)$, which describes the scalar mode (\ref{VectScalBranch}), we encounter two vector modes, and inspired by (\ref{tempVpm}), we write the gauge field as
\bea\label{CfieldZeta0}
C_i dx_i=e^{i n_1\gamma_L+in_2\gamma_R}\left[V_y dy_1+V_+ (d\gamma_L+d\gamma_R)+V_- (d\gamma_L-d\gamma_R)\right].
\eea
Assuming that $\lambda_{vector}\ne 0$, we find the expression for $V_y$ in terms of $(V_+,V_-)$:
\bea\label{Vyeqn}
V_y=-i\frac{(n_1+n_2)(1-y_1)V'_++(n_1-n_2)(1+y_1)V'_-}{n_1^2+n_2^2+\la(y_1^2-1)-2n_1n_2 y_1}
\eea
It turns out that the components $(V_+,V_-)$ decouple only if $n_2=\pm n_1$, so in the general case we 
write 
\bea
V_{+}={\hat V}_{+}+u_1 (n_1^2-n_2^2){\hat V}_{-},\quad
V_{-}={\hat V}_{-}+u_2 (n_1^2-n_2^2){\hat V}_{+}
\eea
with undetermined constants $(u_1,u_2)$. Direct calculations show that equations for 
$({\hat V}_+,{\hat V}_-)$ decouple only for the specific values of $u_{1,2}$:
\bea
u_1=u_2=-\frac{1}{[\sqrt{\la-n_1^2}+\sqrt{\la-n_2^2}]^2}\,,
\eea
and the eigenvalue problems are 
\bea\label{VectODEzeta0}
&&\hskip -1cm \frac{d}{dy_1}\left[\frac{(1-y_1^2){\hat V}_{-}'}{\la y_1-n_1n_2-\mu}\right]+\frac{\la y_1-n_1n_2+\mu}{\la(y_1^2-1)}{\hat V}_{-}=0\nn
\\
&&\hskip -1cm \frac{d}{dy_1}\left[\frac{(1-y_1^2){\hat V}_{+}'}{\la y_1-n_1n_2+\mu}\right]+\frac{\la y_1-n_1n_2-\mu}{\la(y_1^2-1)}{\hat V}_{+}=0\nonumber
\eea
Here we defined 
\bea
\mu=\sqrt{(\la-n_1^2)(\la-n_2^2)}
\eea
Note that the ansatz (\ref{CfieldZeta0}) can be rewritten as
\bea\label{CfieldZeta0v1}
C_i dx_i=e^{i n_1\gamma_L+in_2\gamma_R}\left[V_y dy_1+
q_+({\hat V}_++{\hat V}_-)d\gamma_L+q_-({\hat V}_+-{\hat V}_-)d\gamma_R\right],
\eea
where
\bea
q_+=(1+u_2 n_1^2-u_2 n_2^2),\quad q_-=(1-u_2 n_1^2+u_2 n_2^2).\nonumber
\eea
In particular,
\bea
\frac{q_-}{q_+}=\left[\frac{\la-n_2^2}{\la-n_1^2}\right]^{1/2}\,.
\eea
Similar to \eqref{VpmScalSpecial}, functions $( \hat{V}_+,\hat{V}_-)$ can be written as 
\bea\label{VpmAsB1}
{\hat V}_\pm(y_1)=C_\pm\left[(1-y_1^2)B_1'+\frac{1}{M}
\left[\la y_1-n_1n_2\pm\mu\right]B_1\right]\,,
\eea
where function $B_1$ satisfied the differential equation (\ref{SO4scalODE}), and parameters $(\la,\nu_1,M)$ are related by
\bea\label{EigenValVecZet0Ap}
\la=M^2,\quad \nu_1=M(M+1).
\eea
The two free parameters $C_\pm$ characterize two different degrees of freedom of the vector fields. As demonstrated in section \ref{SecSubSub2}, the expressions for the vector field become especially simple if one considers $C_+= C_-$ or $C_+=-C_-$ (see (\ref{AnstzForSU2})--(\ref{a3aPmCoef}) and (\ref{AnstzForSU2b2})--(\ref{a3aPmCoefb2})). In the next subsection we will extend these two polarizations to the modified equations for the vector field (\ref{temp1Fij}) with a nontrivial value of 
$\zeta$. 

\subsection{Modified vector equation: arbitrary $\zeta$.}
\label{AppSU2sub2}

To demonstrate separation of variables and decoupling of various components in equations (\ref{temp1Fij}) for all values of $\zeta$, we introduce the frames (\ref{LeftFramesMain}), 
\bea\label{LeftPro}
&&e_{\bf 3}^\mu\d_\mu=\d_{\gamma_R},\quad e^\mu_\pm\d_\mu
=-\frac{e^{\mp i\gamma_R}}{2\sqrt{1-y_1^2}}\left[(1-y_1^2)\d_{y_1}\pm i(y_1\d_{\gamma_R}-
\d_{\gamma_L})\right],\\
&&e_3^\mu e_3^\mu+\frac{1}{2}\left(e_+^\mu e_-^\nu+e_+^\nu e_-^\mu\right)=2g^{\mu\nu}.\nonumber
\eea
and impose the ansatz (\ref{AnstzForSU2})\footnote{A similar ansatz for a vector field in a five dimensional black hole with two equal angular momenta was also considered in \cite{ANote}.}
\bea\label{AnstzForSU2App}
e_{\bf 3}^\mu C_\mu=a_3e_{\bf 3}^\mu \d_\mu Z,\quad
e_{\bf +}^\mu C_\mu=a_+e_{\bf +}^\mu \d_\mu Z,\quad
e_{\bf -}^\mu C_\mu=a_-e_{\bf -}^\mu \d_\mu Z\,.
\eea
Here $(a_3,a_+,a_-)$ are undetermined constant coefficients, and function $Z$ is related to the solution of the scalar equation (\ref{SO4scalODE}) by
\bea
Z=B_1 e^{i n_1\gamma_L+in_2\gamma_R}\,.
\eea
Such $Z$ after normalization can be expressed in terms of the Wigner's coefficients,
\bea\label{WignerDefine}
Z= D_{p,q}^J\,,\quad  p=n_2,\quad q=n_1,
\eea 
which obey an important set of identities \footnote{Winger's functions are naturally defined with Euler's angle parametrization and one can check these identities explicitly in that parametrization of the group element. }:
\bea\label{tempFrameDer}
e^\mu_{\bf 3}\d_\mu D_{p,q}^J=ip D_{p,q}^J,\quad e_+^\mu\d_\mu D_{p,q}^J=\frac{1}{2}c_{p} D_{p-1,q}^J,\quad e^\mu_-\p_\mu D_{p,q}^J=-\frac{1}{2}c_{p+1}D_{p+1,q}^J\,.
\eea
Here we defined
\bea
c_{p}=\sqrt{(J+p)(J-p+1)}\,.
\eea
This leads to an alternative form of the ansatz (\ref{AnstzForSU2App}):
\bea\label{SU2AnsAlt}
e_{\bf 3}^\mu C_\mu=a_3\,ip D_{p,q}^J,\quad
e_{\bf +}^\mu C_\mu=a_+\, \frac{c_{p}}{2} D_{p-1,q}^J,\quad
e_{\bf -}^\mu C_\mu=-a_-\,\frac{c_{p+1}}{2}D_{p+1,q}^J\,.
\eea
To evaluate the field strength, we need the expressions for the spin connections in the frames (\ref{AnstzForSU2App}). Although the spin coefficients 
$\Gamma^a_{bc}$,
\bea 
d e^a+\Gamma^a_{cb}e^c\wedge e^b=0,
\eea 
can be easily evaluated from the definition above, it is instructive to compute these coefficients using the group--theoretic analysis. Specifically, substituting the expression for the group element \eqref{ParamSO4} into the definition of the left--invariant forms,
\bea
\sigma^a=\mbox{Tr}(T^a g^{-1}dg)\,.
\eea
we arrive at a very simple relation
\bea
e^a=-i \sigma^a\,.
\eea
Then the Maurer--Cartan equations for the left--invariant forms lead to the expressions for the spin coefficients in terms of the structure constants of $SU(2)$:
\bea
d\sigma^a+\frac{i}{2} f^a_{~bc}\sigma^b\wedge \sigma^c=0\quad\Rightarrow\quad 
\Gamma^a_{bc}=-\frac{1}{2}f^a_{bc}\,.
\eea
These coefficients lead to the expression for the frame components of the modified field strength,
\bea 
\mathcal{F}_{ab}\equiv F_{ab}+\zeta H_{ab}^c C_c=\p_a C_b-\p_b C_a+(f_{ab}^c+\zeta H_{ab}^c) C_c,\quad H_{+-}^3=\frac{i}{4}\,,
\eea
and substitution of the ansatz (\ref{SU2AnsAlt}) and the derivatives (\ref{tempFrameDer}) gives 
\bea\label{FieldStrength}
&&\mathcal{F}_{+-}=\frac{1}{4}\left(-c_{p+1}^2a_{-}+c_{p}^2 a_{+}-2p(1+\zeta) a_3\right)D_{p,q}^J,\\
&&\mathcal{F}_{+3}=\frac{i}{2}c_{p}p \left( a_3- (1+\zeta) a_+\right)D_{p-1,q}^J,\\
&& \mathcal{F}_{-3}=\frac{i}{2}c_{p+1}p\left((1-\zeta) a_--a_3\right)D_{p+1,q}^J.
\eea 

To proceed, we project the equations (\ref{temp1Fij}) to the frames (\ref{AnstzForSU2App}). 
We begin with the derivative of the modified field strength\footnote{In this appendix, we denoting the spacetime indexes with Greek letters, $(\mu,\nu,\dots)$, and the local frame indexes with Latin letters, $(a,b,\dots)$.}:
\bea\label{tempApp}
\nabla_\mu \mathcal{F}^{\alpha\beta} &=& \p_\mu \mathcal{F}^{\alpha\beta}+\Gamma^\alpha_{\mu\lambda}\mathcal{F}^{\lambda\beta}+\Gamma^\lambda_{\mu\beta}\mathcal{F}_{\alpha\lambda}\nn
&=&\p_\mu (\mathcal{F}^{bc}e_b^\alpha e_c^\beta)+\Gamma^\alpha_{\mu\lambda}\mathcal{F}^{bc}e_b^\lambda e_c^\beta+\Gamma^\lambda_{\mu\beta}\mathcal{F}_{bc}e^\alpha_b e^\lambda_c\nn
&=&\p_\mu \mathcal{F}^{bc} e_b^\alpha e_c^\beta+\mathcal{F}^{bc}\left(\nabla_\mu e^\alpha_b e_c^\beta+\nabla_\mu e_c^\beta e_b^\alpha\right).
\eea 
Here we used the relation
\bea 
\Gamma^a_{bc}=e^a_\alpha  e^\beta_b \nabla_\beta e_c^\alpha\,.
\eea 
Substitution of (\ref{tempApp}) into the eigenvalue problem (\ref{temp1Fij}) gives \footnote{The factor $2$ before $\Lambda$ is for our convenience to compare the result here with the results in the $(y,\gamma_L,\gamma_R)$ coordinates.}
\bea 
\nabla_\mu \mathcal{F}^{\mu \beta}+2\Lambda C^\beta &=&\p_a \mathcal{F}^{ac} e_c^\beta+\mathcal{F}^{bc}(\nabla_\alpha e_b^\alpha e_c^\beta+e_b^\alpha \nabla_\alpha e_c^\beta)+2\Lambda C^\beta=0.
\eea
Multiplication by $e_d^\beta$ and summation over $\beta$ 
gives the final form of the system (\ref{temp1Fij}) projected to frames\footnote{Recall that $\Gamma^a_{bc}=-\frac{1}{2}f^a_{bc}$.}:
\bea 
\p_a \mathcal{F}^{ad}-\frac{1}{2} \mathcal{F}^{bd}f^a_{ab}-\frac{1}{2}\mathcal{F}^{bc}f^d_{bc}+2\Lambda C^d=0.
\eea 
More explicitly, we find three independent equations:
\bea\label{tempDec18}
&&\p_- \mathcal{F}_{+3}+\p_+\mathcal{F}_{-3}-i \mathcal{F}_{-+}+\frac{\Lambda}{2}C_3=0,\nn
&&\p_+ \mathcal{F}_{-+}+\frac{1}{2}\p_3 \mathcal{F}_{3+}-\frac{i}{2}\mathcal{F}_{+3}+\frac{\Lambda}{2}C_+=0,\\
&&\p_-\mathcal{F}_{+-}+\frac{1}{2}\p_3 \mathcal{F}_{3-}+\frac{i}{2}\mathcal{F}_{-3}+\frac{\Lambda}{2}C_-=0.\nonumber
\eea 
Substitution of (\ref{SU2AnsAlt}) and (\ref{FieldStrength}) reduces (\ref{tempDec18}) to a system of {\it algebraic} equations for coefficients $(a_3,a_+,a_-)$:
\bea 
&&\hskip -1cm M\left[\begin{array}{c}
a_3\\a_+\\a_-
\end{array}\right]=
\left[\begin{array}{c}
0\\0\\0
\end{array}\right]\,,\\
&&\hskip -1cm
M\equiv\left[
\begin{array}{ccc}
p[c_{p}^2+c_{p+1}^2-2(\Lambda-\zeta-1)]&-(1+p+\zeta)&1-p+\zeta\\
2p(1+p+\zeta)&2\Lambda-2p(p+\zeta)-c_{p}^2&c_{p+1}^2\\
2p(1-p+\zeta)&-c_{p}^2&2p(p-\zeta)-2\Lambda+c_{p+1}^2
\end{array}
\right]\,.\nonumber
\eea 
Setting the determinant of the matrix $M$ to zero,
\bea 
[J(J-\zeta)-\Lambda][(1+J)(1+J+\zeta)-\Lambda]\Lambda=0,
\eea
we find two physical polarizations of the vector field, as well as a pure gauge\footnote{Here we recalled the the definition (\ref{WignerDefine}) to go back from parameter $p$ to $n_2$, which is used in the main body of the paper.}:
\bea\label{LambdaApp}
\Lambda&=&J(J-\zeta):\quad a_\pm=\frac{n_2}{n_2\pm J}a_3\,;\nn
\Lambda&=&(1+J)(1+J+\zeta):\quad a_\pm=\frac{n_2}{n_2\mp (J+1)}a_3\,;\\
\Lambda&=&0:\quad a_-=a_+=a_3\,.\nonumber
\eea
The last line gives $F_{\mu\nu}=0$, but ${\cal F}_{\mu\nu}$ is still nontrivial:
\bea
\mathcal{F}_{+-}=-\frac{p\zeta}{2}a_3 D_{p,q}^J,\quad
\mathcal{F}_{+3}=-\frac{ip\zeta}{2}c_{p} a_3 D_{p-1,q}^J,\quad
\mathcal{F}_{-3}=-\frac{ip\zeta}{2}c_{p+1}a_3 D_{p+1,q}^J.
\eea 
This field is divergence--free, so it gives $\Lambda=0$.

To threat the first two solutions from (\ref{LambdaApp}) in a unified fashion, we observe that 
$Z= D_{n_2,n_1}^J$ is a solution of the scalar equation (\ref{SO4scalODE}) with an eigenvalue
\bea
\nu_1=J(J+1),\qquad J\ge 0. 
\eea
The last expression can be rewritten as 
\bea
\nu_1=M(M+1)\quad \mbox{with}\quad M=J \quad \mbox{or}\quad M=-J-1\,.
\eea
This concluded the rerivation of the vector eigenvalues (\ref{VecEigen}). 
Taking the first option for the first line in (\ref{LambdaApp}) and the second option for the second line, we can summarize both branches as 
\bea\label{LambAppFin}
\Lambda=M(M-\zeta):\quad a_\pm=\frac{n_2}{n_2\pm M}a_3\,\quad \nu_1=M(M+1),
\eea
where $M$ can take both positive and negative values. This concludes the derivation of the vector eigenvalues (\ref{VecEigen}).

Although it is not obvious {\it a priori}, it is straightforward to check that the solutions (\ref{SU2AnsAlt})  (\ref{LambAppFin}) satisfy the ``Lorenz gauge'' condition\footnote{Since we are dealing with a massive vector field $C$, the equations (\ref{temp1Fij}) are not invariant under 
the $C\rightarrow C+df$ transformations.}:
\bea 
\nabla_\mu C^\mu=\nabla_a C^a=2n_2^2 a_3+(J+1+n_2)(J-n_2)a_-+(J+n_2)(J+1-n_2)a_+=0.
\eea 
This fact plays an important role in section \ref{SecVecSO4}, where the $SU(2)$ vectors are combined to produce separable solutions for the vector fields on $SO(4)$. 

This method can be easily generalized to model with higher rank semi-simple groups. Take a highest weight representation in which states can be labelled by their weight $\vec{r}$. The proper ansatz of the vector field will be
\bea \label{GeneralAnsatz}
e_\mu^{H_i}\p^\mu A_\mu=\sum_{\vec{r}}a_{H_i} D_{\vec{r}},\quad e_\mu^{E_{\vec{\alpha}_i}}\p^\mu A_\mu=a_{E_{\vec{\alpha}_i}}D_{\vec{r}-\vec{\alpha}_i},
\eea 
where $H_i$ correspond to the Cartan subgroup (the analogue of $T_3$), $\vec{\alpha}_{i}$ correspond to the roots (the analogue of $T_{\pm}$) and $D_{\vec{r}}$ are the higher dimensional analogue of Wigner's  functions. Using the general commutation relations between the Cartan and root operators, one can see that indeed that different modes will not mix in any components of the field strength if we assume there are no degenenaries. However to derive the corresponding eigenvalue problem is very cumbersome.

\section{Geometries for the gauged WZW models}
\label{AppGWZW}
\renewcommand{\theequation}{B.\arabic{equation}}
\setcounter{equation}{0}
The gauged WZW on coset $G/H$ can be constructed by integrating out the gauge fields $A_\pm$  corresponding to the subgroup $H$ from WZW on $G$. In this appendix, we present the construction of this geometric background.
It is convenient to separate the generators $T^A$ of the group $G$ into $T^a$ corresponding to the subgroup $H$ and $T^\alpha$ corresponding to the coset $G/H$ and define the left and right Maurer-Cartan forms on $G$ as
 \bea\label{Form}
 &&L_M^A=-i\text{Tr}(T^A g^{-1}\p_M g),\quad R_\mu ^A=-i\text{Tr}(T^A \p_M gg^{-1}),\nonumber\\
 && R^A_M =D_{AB}L_M ^B,\quad D_{AB}=\text{Tr}(T_{A}g T_{B}g^{-1}),\quad g\in G.
 \eea 
In terms of the left and right forms the corresponding metric, Kalb-Ramond and dilaton field of the gauged WZW are given by
\bea 
&&G_{\mu\nu}=\frac{k}{2\pi}\left(\eta_{AB}L^A_\mu L^B_\nu-L^a_\mu(D_{ab}-\eta_{ab})^{-1}R_\nu^b-R_a(D_{ab}-\eta_{ab})^{T}L_\nu^b \right)\, ,\\
&&B_{\mu\nu}=\frac{k}{2\pi}\left(B_{\mu\nu}^{0}-L^a_\mu(D_{ab}-\eta_{ab})^{-1}R_\nu^b+R_a(D_{ab}-\eta_{ab})^{T}L_\nu^b\right),\nn
&&e^{-2\Phi}=\mbox{det}(D_{ab}-\eta_{ab}),
\eea 
where $B^0$ is defined by $H^0=dB^0$ with
\bea 
H^0_{MNP}=f_{ABC}L^A_M L^B_N L^C_P,\quad f_{ABC}=-i \mbox{Tr}([T_A,T_B]T_C).
\eea 
After some manipulations, the metric can also be written in terms of local frames as
\bea 
&&G_{\mu\nu}=\frac{k}{2\pi}\,\eta_{\alpha\beta}\,e^\alpha_\mu e^\beta_\nu=\frac{k}{2\pi}\,\eta_{\alpha\beta}\,\hat{e}^\alpha_\mu \hat{e}^\beta_\nu,\\
&&e^\alpha_\mu=L^\alpha_\mu-D^T_{a\alpha}(D_{ab}-\eta_{ab})^{-T}L^b_\mu \label{},\quad \hat{e}^\alpha_\mu=R^\alpha_\mu-D_{\alpha a} (D_{ab}-\eta_{ab})^{-1}R^b_\mu \label{twoframes}.
\eea 
In the group model  \eqref{WZWaction}, there are also two natural local frames $L^A_M$ and $R^A_M$ which are related to the $G_L\times G_R$ isometries of the group model. By gauging the subgroup $H$, the $G_L\times G_R$ isometries are mostly explicitly broken and 
the resulted gWZW model should only depend on the gauge invariant quantities. As a result the two local frames \eqref{twoframes} of gWZW are gauge invariant projections and restrictions of $L^A_M$ and $R^A_M$. Therefore the two local frames \eqref{twoframes} are the ones (after some linear combinations) we used in our separable ansatz \eqref{AnstzForSU2} and \eqref{VectAnstz} inherited from the general form \eqref{GeneralAnsatz}.

\section{Scalar wavefunctions for the $SO(5)$ sigma model}
\renewcommand{\theequation}{C.\arabic{equation}}
\setcounter{equation}{0}

In this appendix we provide some technical details relevant for deriving several classes of scalar eigenfunctions on the background of the $SO(5)$ WZW model. 

\subsection{Solutions with linear $z$--dependence}
\label{AppSO5Linear}

In section \ref{SecSubXlinear} we outlined the procedure for starting with a general expression (\ref{LinearZform}) for the eigenfunction linear in $(X_1,X_2,X_2,X_4)$ and using various {\it algebraic} relations following from the Helmholtz equation (\ref{HelpmSO5}) to derive the options (\ref{HelpmSO5}) that correspond to two types of branches (\ref{LinearZbranches}). In this Appendix, we will begin with (\ref{LinearZbranches}) and derive the expressions for functions ($g_1,g_2,g_3,g_4)$. We will focus on the first option in (\ref{LinearZbranches}), and the second one can be analyzed in the same way. 

Let us define a convenient function
\bea
F=(1+R^2)^q(\nabla^2+\Lambda)\Psi.
\eea
Once the first solution from (\ref{LinearZbranches}),
\bea
n_2=n_1-1,\ n_3=n_4:&&\hskip -0.5cm
\Psi=\frac{E_{n_1,n_2,n_3,n_4}}{(1+R^2)^q}\left[g_1(Y_1,Y_2){\bar z}_1+g_2(Y_1,Y_2){z}_2\right],
\eea
is substituted, function $F$ must vanish. This leads to complicated overdetermined equations for $g_1$ and $g_2$, but one of the necessary conditions is very simple:
\bea
\left.\frac{\d^2 F}{\d X_3^2}\right|_{X_1=i X_2} =0:\quad
\left[(1+Y_2^2)\d_{Y_2}+(1+Y_1^2)\d_{Y_1}\right]g_2=0.
\eea
The most general solution of the resulting first order PDE is 
\bea
g_2=h_2\left[\frac{Y_1-Y_2}{1+Y_1 Y_2}\right],
\eea
where $h_2$ is an arbitrary function of its argument. Next we look at another necessary condition:
\bea
\left.\frac{\d}{\d X_4}\left(F|_{X_3=-i X_4}\right)\right|_{X_4=0}=0:\quad \left[(1+Y_2^2)\d_{Y_2}+(1+Y_1^2)\d_{Y_1}\right]g_1=0
\eea
This gives
\bea
g_1=h_1\left[\frac{Y_1-Y_2}{1+Y_1 Y_2}\right]
\eea
All remaining equations contain $(Y_1,Y_2)$ only in one combination 
\bea
y_-=\frac{Y_1-Y_2}{1+Y_1 Y_2},
\eea
so we find an overdetermined system of {\it ordinary differential equations} for 
$(h_1[y_-],h_2[y_-])$ with coefficients depending on $(X_1,X_2,X_2,X_4)$. One of these equations gives the expression (\ref{h1Branch1}) for $h_1$ in terms of $h_2$, then the consistency of the remaining system for $h_2$ leads to two options for $\Lambda$:
\bea
\Lambda=2q(q+1)\quad\mbox{or}\quad \Lambda=2q(q+2).
\eea
In both cases, the system collapses to a single hypergeometric equation for $h_2$, and the solutions are given by (\ref{LinearZsolnE1}).

The second option in (\ref{LinearZbranches}) can be analyzed in the same way, and the result is 
(\ref{LongBranch2})--(\ref{LinearZsolnE2}).

\subsection{Wavefunctions in the symmetric representations}
\label{AppSO5comb}
The goal of this subsection is to derive the expression (\ref{FinalProdGeom}) from the general result (\ref{tempDec25}). To do so, we focus on the wavefunction (\ref{tempDec25}) containing four building blocks: $(g_{z5},g_{{\bar z}5},g_{w5},g_{{\bar w}5})$. To evaluate the relevant combinatorial factors, it is conveinient to white the leading term in (\ref{tempDec25}) more explicitly as
\bea
g_{a_1 5}\dots g_{a_{2s} 5}\rightarrow [g_{z5}]^{q_1} [g_{w5}]^{q_2} x^{s_1}y^{s_2},
\eea
where we defined
\bea
x=g_{z5}g_{\bar z5},\quad y=g_{w5}g_{\bar w5}
\eea
Furthermore, relations (\ref{CombinKron}) ensure that the only two types of contractions in (\ref{tempDec25}): $g_{z5}$ with $g_{{\bar z}5}$ and $g_{w5}$ with $g_{{\bar w}5}$. Therefore, the index $k$ in (\ref{tempDec25}) can be decomposed as $k=k_1+k_2$ to keep track of the contractions of the first and the second types. With this notation we find
\bea
&&\frac{\delta_k\Pi_{2s-2k}}{ [g_{z5}]^{q_1} [g_{w5}]^{q_2} x^{s_1}y^{s_2}}
=\sum_{p=0}^k\left\{
 2^p p!\left[\frac{(s_1+q_1)!}{p!(s_1+q_1-p)!}\right]\left[\frac{s_1!}{p!(s_1-p)!}\right]\right.\nonumber\\
 &&\qquad\times \left.2^{k-p} (k-p)!\left[\frac{(s_2+q_2)!}{(k-p)!(s_2+q_2-k+p)!}\right]
 \left[\frac{s_2!}{(k-p)!(s_2-k+p)!}\right]
 \frac{1}{x^py^{k-p}}\right\}\nn
 &&=2^k\sum_{p=0}^k\frac{x^{-p}y^{p-k}}{p!(k-p)!}
  \frac{(s_1+q_1)!}{(s_1+q_1-p)!}\frac{s_1!}{(s_1-p)!}\frac{(s_2+q_2)!}{(s_2+q_2-k+p)!}
 \frac{s_2!}{(s_2-k+p)!}
 \eea
 Substitution into (\ref{tempDec25}) gives
 \bea\label{tempApell}
 \Phi&=&\sum_{k=0}^s(-1)^k\frac{(4s+1-2k)!!}{(4s+1)!!}\delta_k\Pi_{2s-2k}\nn
 &=&
 [g_{z5}]^{q_1} [g_{w5}]^{q_2}
 \sum_{k=0}^s(-1)^k\frac{(4s+1-2k)!!}{(4s+1)!!}2^k\nn
 &&\times\sum_{p=0}^k\frac{x^{s_1-p}y^{s_2+p-k}}{p!(k-p)!}
  \frac{(s_1+q_1)!}{(s_1+q_1-p)!}\frac{s_1!}{(s_1-p)!}\frac{(s_2+q_2)!}{(s_2+q_2-k+p)!}
 \frac{s_2!}{(s_2-k+p)!}\nn
 &=& [g_{z5}]^{q_1} [g_{w5}]^{q_2}\sum_{m,n}\frac{x^m y^n}{m!n!(-2)^{m+n-s_1-s_2}}
 \frac{(4s+1+2m+2n-2s_1-2s_2)!!}{(4s+1)!!}
 \nn
 &&\times\frac{(-s_1)_m}{(-1)^m} \frac{(-s_2)_n}{(-1)^n}\frac{(s_1+q_1)!}{(m+q_1)!}\frac{(s_2+q_2)!}{(n+q_2)!}
 \eea
 At the last stage we used the relation 
 \bea\label{FactorialOne}
 (-s_1)_m=(-s_1)(-s_1+1)\dots(-s_1+m-1)=(-1)^m\frac{s_1!}{(s_1-m)!}
 \eea
 for the coefficients $(a)_n=a\dots (a+n-1)$ which appear in the definition of the hypergeometric function and its generalization (\ref{ApellDef}). To simpify the expression (\ref{tempApell}), we observe that
the definition of $s$ from (\ref{tempDec25}) implies $s=s_1+s_2+\frac{q_1+q_2}{2}$. Then
\bea\label{FactorialTwo}
&&(4s+1+2m+2n-2s_1-2s_2)!!=(2[s_1+s_2+q_1+q_2]+2[m+n]+1)!!\nn
&&\qquad= 2^{m+n}(s_1+s_2+q_1+q_2+\frac{3}{2})_{m+n}\,(2[s_1+s_2+q_1+q_2]+1)!!
\eea
Substituting this into (\ref{tempApell}) and dropping a complicated but irrelevant overall factor which depends on $(s_1,s_2,q_1,q_2)$, we find\footnote{To arrive at this expression we also used the 
relations $(m+q_1)!=(q_1+1)_m q_1!$ and $(n+q_2)!=(q_2+1)_n q_2!$.}
\bea\label{temAppell2}
 \Phi
 &\propto& [g_{z5}]^{q_1} [g_{w5}]^{q_2}\sum_{m,n}\frac{x^m y^n}{m!n!}
 \frac{(s_1+s_2+q_1+q_2+\frac{3}{2})_{m+n}}{(q_1+1)_m(q_2+1)_n}(-s_1)_m(-s_2)_n\nn
 &\propto&[g_{z5}]^{q_1} [g_{w5}]^{q_2}F_2\left[\frac{3}{2}+q_1+q_2+s_1+s_2,-s_1,-s_2;q_1+1,q_2+1;x,y\right].
 \eea
Here we used the definiton (\ref{ApellDef}) of the Appell series. Equation (\ref{temAppell2}) completes the derivation of the wavefunction (\ref{FinalProdGeom}). 

\bigskip

We conclude this subsection by justifying the wavefunction (\ref{eqn146extra}) for the special situation of $s_2=0$. In this case the Appell series reduce to the standard hypergeometric function, but one can also easily derive the result (\ref{eqn146extra}) from the first principles without appealing to the reduction formulas. For the solution with $s_2=0$, the contractions can happen only between $g_{z5}$ and 
$g_{\bar z 5}$, so the contraction rule becomes
\bea
&&\delta_p\Pi_{2s-2p}
=
 2^p p!\left[\frac{(s_1+q_1)!}{p!(s_1+q_1-p)!}\right]\left[\frac{s_1!}{p!(s_1-p)!}\right]
x^{s_1-p}(g_{z5})^{q_1}(g_{w5})^{q_2}
 \eea
 Then the summation (\ref{tempDec25}) gives
\bea
\Phi&=&(g_{z5})^{q_1}(g_{w5})^{q_2}
\sum_{k}(-2)^k\frac{(4s+1-2k)!!}{(4s+1)!!} \frac{x^{s_1-k}}{k!}
\frac{(s_1+q_1)!}{(s_1+q_1-k)!}\frac{s_1!}{(s_1-k)!}
\nn
&\propto&(g_{z5})^{q_1}(g_{w5})^{q_2}
\sum_{n}\frac{1}{(-2)^n}\frac{(2[s_1+q_1+q_2]+2n+1)!!}{(4s+1)!!} \frac{x^{n}}{(s_1-n)!}
\frac{(s_1+q_1)!}{(n+q_1)!}
\frac{s_1!}{n!}\nn
&\propto&(g_{z5})^{q_1}(g_{w5})^{q_2}
\sum_{n}\frac{(s_1+q_1+q_2+\frac{3}{2})_n(-s_1)_n}{(n+q_1)!} \frac{x^{n}}{n!}\nonumber
\eea
We used the expressions (\ref{FactorialOne}) and (\ref{FactorialTwo}) for the factorials and double factorials in terms of the gamma functions, and, as before, we dropped the complicated but irrelevant constant prefactors in front of $\Phi$. Performing the sum in the last expression, we arrive at the final result (\ref{eqn146extra}):
\bea
\Phi=(g_{w5})^{q_2}(g_{z5})^{q_1}F[-k,\frac{3}{2}+k+q_1+q_2;1+q_1;g_{z5}g_{\bar z 5}],\quad s=q_1+q_2+2k.
\eea
This concludes our justification of various combinatorial formulas used in section \ref{SecSubSymRep}.

\end{document}